\documentclass[journal] {IEEEtran}
\usepackage{tikz}
\usetikzlibrary{arrows,positioning,fit,shapes,calc,decorations.pathreplacing,matrix,patterns}
\usepackage{amsmath}
\usepackage[caption=false,font=footnotesize,labelfont=rm,textfont=rm]{subfig}
\usepackage{stfloats}
\usepackage{lineno,hyperref}
\usepackage{xcolor}
\usepackage{extarrows}
\usepackage{algorithm}
\usepackage{algorithmic}
\usepackage{amssymb}
\usepackage{graphicx}
\usepackage[justification=centering]{caption}
\usepackage{cite}
\usepackage{enumitem}

\usepackage{stfloats}
\usepackage{amsthm}
\usepackage{geometry}
\usepackage{booktabs} 
\usepackage{cuted}

\newcommand{\ba}{\boldsymbol{a}}

\newcommand{\bd}{\boldsymbol{d}}
\newcommand{\bx}{\boldsymbol{x}}
\newcommand{\by}{\boldsymbol{y}}
\newcommand{\bh}{\boldsymbol{h}}
\newcommand{\bw}{\boldsymbol{w}}
\newcommand{\bz}{\boldsymbol{z}}
\newcommand{\bu}{\boldsymbol{u}}
\newcommand{\bp}{\boldsymbol{p}}
\newcommand{\bs}{\boldsymbol{s}}
\newcommand{\bq}{\boldsymbol{q}}

\newcommand{\be}{\boldsymbol{e}}

\newcommand{\btheta}{\boldsymbol{\theta}}
\newcommand{\bvarphi}{\boldsymbol{\varphi}}
\newcommand{\bDelta}{\boldsymbol{\Delta}}

\newcommand{\bE}{\boldsymbol{E}}

\newcommand{\bA}{\boldsymbol{A}}
\newcommand{\bB}{\boldsymbol{B}}

\newcommand{\bI}{\boldsymbol{I}}

\newcommand{\bD}{\boldsymbol{D}}

\newcommand{\bH}{\boldsymbol{H}}

\newcommand{\bJ}{\boldsymbol{J}}
\newcommand{\bPi}{\boldsymbol{\Pi}}

\newcommand{\bGamma}{\boldsymbol{\Gamma}}
\newcommand{\bLambda}{\boldsymbol{\Lambda}}

\newcommand{\N}{\mathcal{N}}

\newcommand{\mCN}{\mathcal{CN}}
\newcommand{\mO}{\mathcal{O}}
\newcommand{\mA}{\mathcal{A}}

\newcommand{\bmu}{\boldsymbol{\mu}}
\newcommand{\bnu}{\boldsymbol{\nu}}
\newcommand{\cev}[1]{\reflectbox{\ensuremath{\vec{\reflectbox{\ensuremath{#1}}}}}}
\newcommand{\mF}{\mathcal{F}}

\newcommand{\herm}{^{\textrm{H}}}
\newcommand{\tra}{^{\textrm{T}}}

\newcommand{\bbeta}{\boldsymbol{\beta}}

\newcommand{\bgamma}{\boldsymbol{\gamma}}
\newcommand{\bphi}{\boldsymbol{\phi}}

\usetikzlibrary{arrows,positioning,fit,shapes,calc,decorations.pathreplacing,matrix,patterns}

\newgeometry{left=1.5cm, right=1.5cm, top=1.5cm, bottom=1cm}
\tikzstyle{factornode} = [draw, fill=white, circle, inner sep=1pt, minimum size=1.4cm]
\tikzstyle{funnode} = [draw, rectangle,fill=black!100, minimum size = 8mm]

\begin{document} 
	\title{ Message-Passing-Based Bayesian Joint Channel Estimation and Data Detection for Pilot-Guard-Free AFDM  }  
\author{Zhengdao~Yuan, Qinghua Guo, \IEEEmembership {Senior Member, IEEE},  Ming Jin, \IEEEmembership {Senior Member, IEEE}, Zilong Liu, \IEEEmembership {Senior Member, IEEE}, 
and Pei Xiao, \IEEEmembership {Senior Member, IEEE}
\thanks{Z. Yuan is with the Artificial Intelligence Technology Engineering Research Center, Henan Open University,
Zhengzhou 450002, China. He was with the School of  Engineering, University of Wollongong, Wollongong, NSW 2522, Australia (e-mail: yuan\_zhengdao@foxmail.com).}
\thanks{Q. Guo is with the School of Engineering, University of Wollongong, Wollongong, NSW 2522, Australia  (e-mail: qguo@uow.edu.au).}
\thanks{ M. Jin is with the School of Artificial Intelligence, Ningbo University, Ningbo 315211, China (e-mail: jinming@nbu.edu.cn).}
\thanks {Z. Liu is with the School of Computer Science and Electronics Engineering, the University of Essex,
Colchester, CO4 3SQ, United Kingdom (email: zilong.liu@essex.ac.uk).}
\thanks {P. Xiao is with the Institute for Communication
Systems (ICS), Home of the 5GIC \& 6GIC, University of Surrey, United Kingdom (email: p.xiao@surrey.ac.uk).}
}

\maketitle

\begin{abstract}
Existing channel estimation methods for affine frequency division multiplexing (AFDM) over doubly dispersive channels commonly isolate pilots and data with {pilot guard symbols}, resulting in substantial transmission overhead, while fractional Doppler introduces off-grid mismatch and complicates accurate channel estimation. This paper develops a pilot-guard-free AFDM transmission and reception scheme with a Bayesian receiver for joint channel estimation and data symbol detection. Uniformly interleaved pilots are embedded among data symbols without dedicated guard symbols, and the resulting pilot–data coupling is explicitly incorporated into the joint inference problem rather than suppressed through guard-based isolation.   
The channel is modeled using a parametric sparse representation over candidate delay–integer-Doppler components, and its nonlinear dependence on fractional Doppler is handled through successive linearization. Building on a  factor graph representation, an iterative message-passing algorithm is developed to jointly infer path activity, channel gains, fractional-Doppler offsets, noise precision, and data symbols. 
Numerical results demonstrate substantially improved data-transmission efficiency and more accurate channel and fractional-Doppler estimation than pilot-guard-based SBL schemes, together with improved BER performance approaching the ideal-CSI benchmark.
\end{abstract}

\begin{IEEEkeywords}
Affine frequency division multiplexing (AFDM), joint channel estimation and data detection, Bayesian inference, sparse Bayesian learning,  message passing, fractional Doppler.
\end{IEEEkeywords}

\section{Introduction}
\IEEEPARstart{E}merging 6G applications require highly reliable communications in high-mobility environments, where multipath propagation and Doppler shifts/spread give rise to doubly dispersive channels. Under such conditions, Doppler-induced intercarrier interference can severely degrade conventional orthogonal frequency division multiplexing (OFDM). Delay–Doppler-domain waveforms, such as orthogonal time frequency space (OTFS), have attracted considerable attention for their ability to exploit channel diversity in rapidly time-varying environments \cite{raviteja2018interference, deng2025unifying, ge2021receiver,deng2023jamming}. More recently, affine frequency division multiplexing (AFDM) has 
emerged as a promising alternative, employing the discrete affine Fourier transform (DAFT) and appropriately designed chirp parameters to modulate the data symbols onto  multiple orthogonal chirp subcarriers 
\cite{bemani2023affine, rou2024orthogonal, ranasinghe2025joint, li2025affine}. With appropriate chirp parameter design, AFDM can achieve full diversity over doubly dispersive channels and provide performance comparable to OTFS, making it well suited to high-mobility wireless communications \cite{bemani2023affine, ma2023characteristics, luo2025novel}. Moreover, it was shown in \cite{zla} that AFDM is particularly resilient to various types of hardware impairments, such as phase noise, carrier frequency offsets, IQ imbalance, nonlinear power amplifier, and DC offset. Importantly, AFDM can be efficiently implemented through minimum modification of an OFDM system, thus exhibiting excellent backwards-compatibility to legacy wireless infrastructure \cite{zliu}. AFDM has also been adopted as the foundational waveform to support generalized spatial modulation \cite{Sui2026Generalized}, sparse code multiple access (SCMA) \cite{Luo2024AFDM}, and integrated sensing and communication \cite{zlc
}, demonstrating its potential as a versatile waveform candidate. 
Non-orthogonal AFDM has further been investigated to improve spectral efficiency through controllable subcarrier overlapping \cite{Yi2026Non}.

Despite its potential, practical AFDM reception depends critically on accurate channel state information, although a majority of the existing literature assumes perfect CSI \cite{li2026ep, wu2024mpdetect}. In doubly dispersive channels, the time–frequency channel response may vary rapidly, while the underlying physical parameters including the path gains, delays, and Doppler shifts, can remain approximately constant over an AFDM frame \cite{ma2023characteristics}. Such a parametric structure is therefore attractive for channel estimation. Nevertheless, fractional Doppler shifts cause off-grid energy leakage in the discrete affine Fourier domain, spreading each path over multiple channel coefficients rather than a single nominal location. Together with the inherent interference between pilot and data symbols, this leakage makes accurate and spectrally efficient channel estimation challenging \cite{tao2025afdm}.

Existing AFDM channel-estimation methods commonly employ embedded pilots surrounded by guard symbols to suppress pilot–data interference \cite{bemani2023affine, yin2024diagonally, yin2022pilot}. Under this pilot-guard structure, threshold-based path detection offers low computational complexity but is sensitive to noise and is mainly applicable to integer-Doppler channels \cite{yin2022pilot}, whereas approximate maximum-likelihood estimation can accommodate fractional Doppler at the cost of an exhaustive search over candidate delay–Doppler parameters \cite{bemani2023affine}. Lower-complexity alternatives, including linear minimum mean squared error (LMMSE), diagonally reconstructed, and matched-filtering-based channel-estimation schemes, reduce the computational burden but may incur performance degradation relative to more sophisticated estimation methods \cite{zheng2025channel, yin2024diagonally, Li2026Matched}. Recently, sparse Bayesian learning (SBL) has been employed to exploit the inherent sparsity of doubly dispersive channels \cite{wei2022otfsoffgrid, Yang2024AFDM_OffGrid}. In particular, grid-refinement SBL (GR-SBL) and grid-evolution SBL (GE-SBL) mitigate the off-grid mismatch caused by fractional Doppler through iterative grid adaptation, thereby improving channel-parameter estimation accuracy \cite{li2025gridevolution, li2026low}. Nevertheless, these methods still rely on guard symbols, whose overhead grows with the channel delay and Doppler spreads and can substantially reduce spectral efficiency, particularly for short AFDM frames.

To reduce pilot overhead, superimposed-pilot AFDM schemes superimpose pilot symbols onto data symbols, thereby avoiding dedicated pilot guard symbols; however, existing designs are sensitive to noise and have mainly been developed for integer-Doppler channels \cite{zheng2025channel}. More recently, joint channel-estimation and data-detection approaches have been investigated for AFDM, including oversampled receivers and joint channel, data, and parameter estimation frameworks \cite{ranasinghe2025joint, ranasinghe2025joint2}. For MIMO-AFDM, joint sparse-graph receivers integrating detection and decoding have also been developed \cite{Luo2026Joint}. The key idea of joint channel-estimation and data-detection is to exploit unknown data symbols  as additional information for channel estimation, rather than being treated solely as interference. However, existing approaches do not jointly address pilot-guard-free transmission, sparse multipath recovery, continuous fractional-Doppler estimation, and soft symbol detection within a unified low-complexity Bayesian receiver. This motivates the development of a joint receiver that improves spectral efficiency without compromising channel-estimation and detection performance.


Eliminating pilot guard symbols, however, couples channel estimation and symbol detection. Because the pilot and data symbols contribute jointly to the received DAFT-domain observations, the unknown data symbols can no longer be treated independently of channel estimation. At the same time, only a small subset of the candidate delay–integer-Doppler components is active, requiring the receiver to identify the sparse multipath support and estimate the corresponding complex path gains. This inference is further complicated by the nonlinear dependence of the equivalent channel matrix on the unknown fractional-Doppler offsets. To address these coupled challenges, we formulate the received signal as a parametric sparse Bayesian model over candidate delay–Doppler components and locally linearize the fractional-Doppler-dependent channel response using a first-order Taylor expansion. Auxiliary variables are then introduced to decouple the resulting linear mixing and bilinear relationships, yielding a structured factor graph through which the path activity, channel gains, fractional-Doppler offsets, transmitted symbols, and noise precision can be iteratively inferred within a unified Bayesian receiver. The main contributions of this work are summarized as follows:
\begin{enumerate}
\item We develop a pilot-guard-free AFDM transmission and reception scheme in which uniformly interleaved pilots are embedded among data symbols without dedicated guard symbols. Rather than isolating the pilot observations from data interference, the resulting pilot–data coupling is explicitly incorporated into the joint channel-estimation and symbol-detection problem. This eliminates the substantial guard overhead required by conventional embedded-pilot AFDM schemes and improves data-transmission efficiency.
\item We develop a sparse parametric representation of the fractional-Doppler AFDM channel over candidate delay–integer-Doppler components, while retaining the fractional-Doppler offsets as continuous parameters. Their nonlinear effects on the DAFT-domain channel response are handled by successive first-order linearization around the current estimates, enabling iterative refinement of the fractional-Doppler offsets without exhaustive multidimensional search.
\item  We construct a structured factor graph that captures the linear mixing and bilinear coupling among the data symbols, channel gains, and fractional-Doppler parameters. Based on this factorization, an iterative message-passing receiver is developed to jointly infer path activity, channel gains, fractional-Doppler offsets, data symbols, and noise precision. The computational complexity of the proposed receiver is analysed, and extensive simulations are conducted to assess its data-transmission efficiency, channel and fractional-Doppler estimation accuracy, convergence behaviour, and BER performance under different system and channel conditions.


\end{enumerate}


The rest of this paper is organized as follows. Section II presents the AFDM system model and the corresponding doubly dispersive channel model. Section III formulates the joint channel-estimation and symbol-detection problem under the uniformly interleaved pilot structure without pilot guard symbols. Section IV develops the proposed Bayesian message-passing receiver based on fractional-Doppler linearization and sparse channel modelling. Section V evaluates the performance and computational complexity of the proposed method through numerical simulations. Finally, Section VI concludes the paper.

\textit{{Notations}}: Boldface lower-case and upper-case letters denote vectors and matrices, respectively. 
A Gaussian distribution of $x$ with mean {$\hat x$} and variance $v$ is represented by $\N(x;{\hat x}, v)$, and $diag(\ba)$ returns a diagonal matrix with $\ba$ on its diagonal. We use $\bA\cdot\bB$ and $\bA\cdot/\bB$ to denote the element-wise product and division between $\bA$ and $\bB$, respectively. The notation $|\bA|^{.2}$ denotes an element-wise magnitude squared operation for $\bA$. We use $\textbf{1}$, $\textbf{0}$ and $\bI$ to denote an all-one matrix, all-zero matrix and identity matrix with a proper size, respectively. {The notation $x \sim \text{U}[a,b]$ denotes that $x$ has a uniform distribution over $a$ and $b$, and symbol $j$ represents the imaginary unit.}  
\section{AFDM System Model}
\subsection{AFDM Transmitter and Pilot Structure}
\begin{figure}[!t]
\centering
\includegraphics[width=0.48\textwidth]{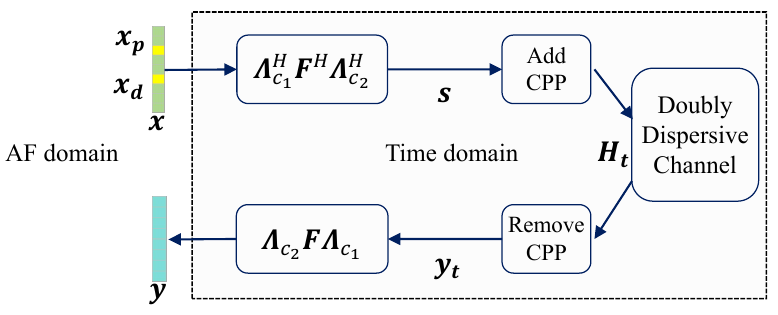}
\centering
\caption{Schematic diagram of the AFDM  system. } 
\label{fig:FigAFDMsys}
\end{figure}

Consider an AFDM system with $N$ subcarriers, as shown in Fig. \ref{fig:FigAFDMsys}. Information bits are mapped to QAM symbols, generating the data vector $\bx_d \in \mathbb{C}^{(N-N_p) \times 1}$. Unlike conventional embedded-pilot AFDM schemes where pilot guard symbols are inserted between the pilot and data symbols, the pilot vector $\bx_p \in \mathbb{C}^{N_p \times 1}$ is uniformly interleaved with the data symbols to construct the transmit vector $\bx \in \mathbb{C}^{N \times 1}$ in the discrete affine Fourier transform (DAFT) domain. 
After serial-to-parallel (S/P) conversion, $\bx$ is transformed to the time domain via inverse DAFT, yielding
\begin{eqnarray}
\bs = \bA\herm \bx \in \mathbb{C}^{N \times 1},
\end{eqnarray}
where $\bs$ denotes the time-domain transmitted signal, and $\bA \in \mathbb{C}^{N \times N}$ represents the DAFT matrix, whose element in the $m$-th row and $n$-th column is given by
\begin{eqnarray}
\bA[m,n] = \frac{1}{\sqrt{N}} \exp\left(-j2\pi\left(c_2 m^2 + \frac{mn}{N} + c_1 n^2\right)\right),
\end{eqnarray}
where $c_1$ and $c_2$ are DAFT modulation parameters. The parameter $c_1$ primarily determines the separation of different delay–Doppler path responses in the DAFT domain, whereas $c_2$ controls the associated phase characteristics and can be appropriately selected to reduce the correlation among the effective channel responses.
Equivalently, the DAFT matrix can also be expressed in a decomposed form $\bA = \bLambda_{c_2} \boldsymbol{F} \bLambda_{c_1}$, where $\boldsymbol{F}$ denotes the normalized $N$-point DFT matrix, and $\bLambda_{c_1}$ and $\bLambda_{c_2}$ are diagonal matrices with the form 
\begin{eqnarray}
\bLambda_c = \mathrm{diag}\left(\exp\left(-j2\pi c n^2\right), n=0,\dots,N-1\right) \in \mathbb{C}^{N \times N}.  \nonumber
\end{eqnarray}
This decomposition shows that the DAFT can be implemented by two chirp-phase operations surrounding a conventional DFT.

To accommodate the multipath delay spread, a chirp cyclic prefix (CPP) of length $L_{cpp}>l_{\text{max}}$ is added to each AFDM block, where $l_{\text{max}}$ denotes the maximum channel delay. The CPP samples are generated as \cite{bemani2023affine}
\begin{eqnarray}
s_{cpp}[n] = s[n+N] \exp\left(-j2\pi c_1\left(N^2 + 2Nn\right)\right), 
\end{eqnarray}
where $n=-L_{cpp},\dots,-1$. After CPP insertion, the resulting time-domain block is converted into serial form and transmitted through the doubly dispersive channel as shown in Fig. \ref{fig:FigAFDMsys}.

\subsection{Doubly Dispersive Channel and Time-Domain Input-Output Model}
Consider a doubly dispersive multipath channel with $P$ propagation paths. 
The discrete time-varying channel impulse response is modeled as
\begin{equation}
g_n[l]
=
\sum_{i=1}^{P}
h_i e^{-j2\pi f_i n}\delta(l-l_i),
\end{equation}
where $P$ is the number of multipath components, and $n$ and $l$ denote the discrete time and delay indices, respectively. 
For the $i$-th path, $h_i$ denotes the complex channel gain, $l_i$ is the 
integer-valued propagation delay, and $f_i$ denotes the normalized Doppler 
frequency with respect to the sampling frequency.
As in \cite{bemani2023affine,li2026low}, the Doppler parameter of the $i$-th path is 
expressed as
\begin{equation}
\nu_i
=
Nf_i
\triangleq
\alpha_i+\beta_i,
\label{eq:doppler_decomposition}
\end{equation}
where $\nu_i$ denotes the normalized Doppler shift with respect to the 
subcarrier spacing, $\alpha_i$ and $\beta_i$ denote its integer and 
fractional components, respectively. In particular, 
$\beta_i\in[-0.5,0.5]$ characterizes the off-grid Doppler offset that 
gives rise to Doppler leakage in the equivalent DAFT-domain channel.

After transmission through the doubly dispersive channel, the received 
time-domain signal is given by
\begin{equation}
y_t[n]
=
\sum_{l=0}^\infty s[n-l]g_n[l]+w[n],
\end{equation}
where $w[n]$ denotes additive Gaussian noise. After removing the CPP, the 
received AFDM block can be written in vector form as
\begin{equation}
\by_t
=
\bH_t\bs+\bw_t,
\end{equation}
where $\by_t\in\mathbb{C}^{N\times 1}$ and 
$\bw_t\in\mathbb{C}^{N\times 1}$ denote the received signal and noise 
vectors, respectively, and $\bH_t\in\mathbb{C}^{N\times N}$ is the 
equivalent time-domain channel matrix. The channel matrix can be expressed as
\begin{equation}
\bH_t
=
\sum_{i=1}^{P}
h_i
\bGamma_i
\bDelta_i
\boldsymbol{\Pi}^{l_i},
\label{eq:time_channel_matrix}
\end{equation}
where $\boldsymbol{\Pi}^{l_i}$ denotes the cyclic delay-shift matrix associated with 
the $i$-th path, such that
{\begin{equation}
\left[\boldsymbol{\Pi}^{l_i}\bs\right]_n
=
s\!\big[(n-l_i)_{\bmod N}\big].
\end{equation}}
The Doppler matrix $\bDelta_i$ is diagonal, with
\begin{equation}
[\bDelta_i]_{n,n}
=
e^{-j2\pi f_i n},
\qquad
n=0,\ldots,N-1.
\end{equation}
The diagonal matrix $\bGamma_i$ accounts for the phase rotation introduced 
when delayed samples originate from the CPP, whose diagonal elements are 
given by
\begin{equation}
[\bGamma_i]_{n,n}
=
\begin{cases}
\exp\!\left[
-j2\pi c_1
\left(
N^2+2N(n-l_i)
\right)
\right],
& 0\leq n<l_i,\\[1mm]
1,
& l_i\leq n\leq N-1.
\end{cases}
\end{equation}
Therefore, \eqref{eq:time_channel_matrix} explicitly captures the effects 
of the complex path gain, CPP-induced phase rotation, Doppler shift, and 
propagation delay of each multipath component.

\subsection{DAFT-Domain Channel Model and Fractional-Doppler Characteristics}

\begin{figure*}[!t]
\centering
\includegraphics[width=0.75\textwidth]{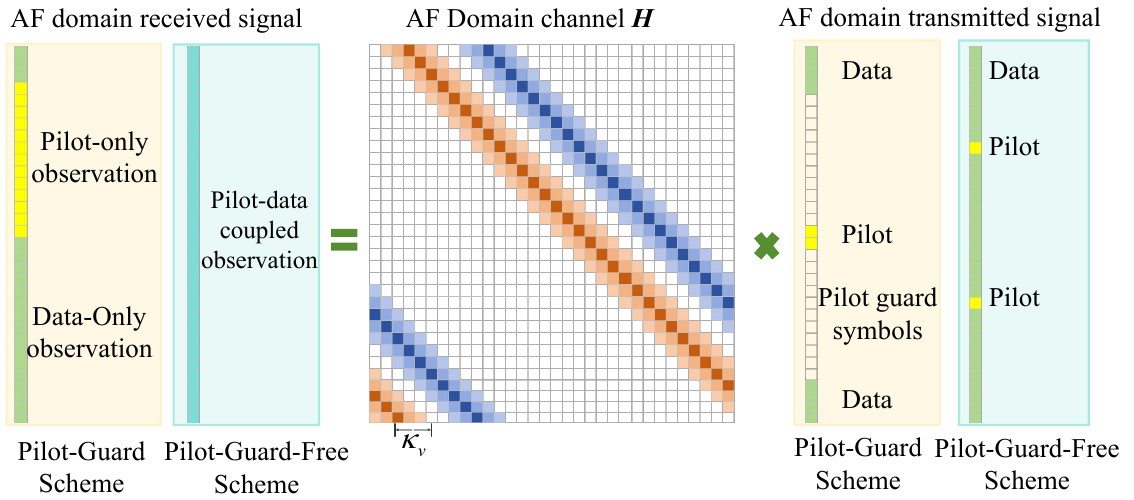}
\centering
\caption{Comparison of transmit and receive signal structures for pilot-guard and pilot guard-free schemes.}
\label{fig:FigChnlH}
\end{figure*}

At the receiver, after CPP removal, the time-domain signal is transformed
into the DAFT domain as
\begin{align}
\by
&=
\bA \bH_t \bA^{H}\bx+\bw \nonumber\\
&=
\sum_{i=1}^{P}
h_i
\underbrace{
\bA\bGamma_i\bDelta_i\boldsymbol{\Pi}^{l_i}\bA^{H}
}_{\triangleq\,\bH_i}
\bx+\bw \nonumber\\
&=
\sum_{i=1}^{P}h_i\bH_i\bx+\bw \nonumber \\
&=
\bH\bx+\bw,
\label{eq:daft_received_signal}
\end{align}
where
\begin{equation}
\bH
=
\sum_{i=1}^{P}h_i\bH_i
\end{equation}
denotes the equivalent DAFT-domain channel matrix, and
$\bw=\bA\bw_t$. Since $\bA$ is unitary, $\bw$ has the same distribution as the time-domain additive Gaussian noise $\bw_t$. For the $i$-th path, the $(p,q)$-th entry of $\bH_i$ can be expressed as
\begin{equation}
H_i[p,q]
=
\frac{\mathcal{F}_i(p,q)}{N}
\exp\!\left\{
j\frac{2\pi}{N}
\left[
Nc_1l_i^2
-q l_i
+Nc_2(q^2-p^2)
\right]
\right\},
\label{eq:path_channel_element}
\end{equation}
where
\begin{align}
\mathcal{F}_i(p,q)
&=
\sum_{n=0}^{N-1}
\exp\!\left[
-j\frac{2\pi}{N}
\left(
p-q+\alpha_i+\beta_i+2Nc_1l_i
\right)n
\right] \nonumber\\
&=
\frac{
\exp\!\left[
-j2\pi
\left(
p-q+\alpha_i+\beta_i+2Nc_1l_i
\right)
\right]-1
}{
\exp\!\left[
-j\frac{2\pi}{N}
\left(
p-q+\alpha_i+\beta_i+2Nc_1l_i
\right)
\right]-1
}.
\label{eq:Fi}
\end{align} Define the nominal DAFT-domain location associated with the $i$-th path as
\begin{equation}
\ell_i^{\mathrm{loc}}
=
\alpha_i+2Nc_1l_i .
\label{eq:path_location}
\end{equation}
When $\beta_i=0$, the path response is concentrated at
\begin{equation}
q
=
\big[p+\ell_i^{\mathrm{loc}}\big]_N,
\end{equation}
where $[\cdot]_N$ denotes the modulo-$N$ operation. Hence, under integer
Doppler, each physical path produces a sharply localized response in the
DAFT domain. When $\beta_i\neq0$, however, the fractional-Doppler component causes
energy leakage into neighboring DAFT-domain coefficients. Specifically,
$|{\mF}_i(p,q)|$ attains its dominant magnitude around
$q=[p+\ell_i^{\mathrm{loc}}]_N$ and decays as $q$ moves away from this
location. Consequently, although $\bH_i$ is generally nonzero over the
entire matrix under fractional Doppler, most of its energy is concentrated
within a narrow band around the nominal path location. Following
\cite{bemani2023affine}, this structure can be approximated by retaining
coefficients within a half-bandwidth $\kappa_v$, resulting in
approximately $2\kappa_v+1$ significant entries per row. {Fig.~\ref{fig:FigChnlH}
illustrates this banded DAFT-domain structure for a representative two-path channel. }

The above DAFT-domain channel structure leads to two important observations
for the receiver design:
\begin{itemize}
\item \textit{Fractional-Doppler-induced spreading:}
Fractional Doppler destroys the strict localization of each path in the
DAFT domain. Although the resulting path matrix is generally dense, most
of its energy remains concentrated around the nominal delay-Doppler
location, yielding an approximately banded structure with
$2\kappa_v+1$ dominant entries. This observation motivates the
parametric fractional-Doppler modeling developed in the next section.
\item \textit{Pilot--data coupling and spectral overhead:}
The spreading of each path creates coupling between pilot and data symbols. {As shown in Fig.~\ref{fig:FigChnlH}},
conventional channel-estimation schemes avoid this coupling by placing
pilot guard symbols around the embedded pilots, such that a subset of
the received samples depends only on known pilot symbols. The required
number of guard symbols is
\begin{equation}
2Q
=
2(l_{\max}+1)
\left(2\alpha_{\max}+2\kappa_v+1\right). \label{eq:Q}
\end{equation}
For example, under the 5G urban channel environment considered in the
manuscript, setting $l_{\max}=10$, $\alpha_{\max}=0$, and $\kappa_v=5$
yields $2Q=242$ pilot guard symbols. When $N=512$, these guard
symbols alone occupy approximately $47\%$ of the DAFT-domain block,
resulting in substantial spectral overhead. To avoid this loss, we
uniformly interleave the pilot vector $\bx_p$ with the data symbols
without pilot guard symbols, as illustrated in {Fig.~\ref{fig:FigChnlH}.} Rather
than eliminating the resulting pilot--data coupling, the proposed receiver
explicitly exploits it to jointly perform channel estimation and symbol
detection.
\end{itemize}

\section{Problem Formulation for Joint Channel Estimation and Symbol Detection and Factor Graph Representation}
\subsection{Sparse Parametric Modeling and Fractional-Doppler Linearization}
From \eqref{eq:daft_received_signal}, the received DAFT-domain signal can
be expressed in terms of the $P$ physical propagation paths as
\begin{equation}
\by
=
\sum_{i=1}^{P}
h_i
\bH(l_i,\alpha_i,\beta_i)\bx
+
\bw,
\label{eq:physical_channel_model}
\end{equation}
where the transmitted data symbols, path gains, delays, and Doppler
parameters are generally unknown at the receiver. Direct estimation
of these quantities is challenging because they are coupled with each other and the equivalent channel matrix depends nonlinearly on the fractional
Doppler parameters.
To circumvent this problem, we consider a
set of $K$ candidate delay--integer-Doppler pairs,
\begin{equation}
\mathcal{G}
=
\left\{
(l_k,\alpha_k)
\right\}_{k=1}^{K},
\end{equation}
where $l_k\in[l_{\min},l_{\max}]$ and
$\alpha_k\in[-\alpha_{\max},\alpha_{\max}]$.
For each candidate pair $(l_k,\alpha_k)$, let $h_k$ denote the
corresponding complex path gain and $\beta_k\in[-0.5,0.5]$ its
fractional-Doppler offset. If the $k$-th candidate path is inactive,
$h_k=0$. Hence, the vector
\begin{equation}
\bh
=
[h_1,\ldots,h_K]^{T}
\end{equation}
is sparse, with its nonzero entries identifying the active delay--Doppler
components. The received signal can therefore be rewritten as
\begin{equation}
\by
=
\sum_{k=1}^{K}
h_k\bH_k(\beta_k)\bx
+
\bw,
\label{eq:sparse_parametric_model}
\end{equation}
where \begin{equation}
\bH_k(\beta_k)
\triangleq
\bA\bGamma_k\bDelta_k(\beta_k)
\bPi^{l_k}\bA^{H}.
\label{eq:candidate_path_matrix}
\end{equation} 
The unknown quantities in \eqref{eq:sparse_parametric_model} are thus
the data symbols, the sparse path gains, and the fractional-Doppler 
offsets associated with the active candidate paths.

The dependence of $\bH_k(\beta_k)$ on $\beta_k$ is nonlinear, making its estimation difficult. To handle this, we use an iterative estimation approach, and linearize $\bH_k(\beta_k)$ at the fractional-Doppler
estimate $\hat{\beta}_k^{(t)}$ from the prior iteration $t$. A first-order Taylor
expansion gives
\begin{align}
\bH_k(\beta_k)
&\approx
\bH_k^{(t)}
+
\left(
\beta_k-\hat{\beta}_k^{(t)}
\right)
\bE_k^{(t)}
\nonumber\\
&=
\bD_k^{(t)}
+
\beta_k\bE_k^{(t)},
\label{eq:taylor_channel}
\end{align}
where
\begin{equation}
\bH_k^{(t)}
\triangleq
\bH_k\!\left(\hat{\beta}_k^{(t)}\right),
\end{equation}
\begin{equation}
\bE_k^{(t)}
\triangleq
\left.
\frac{\partial\bH_k(\beta_k)}
{\partial\beta_k}
\right|_{\beta_k=\hat{\beta}_k^{(t)}},
\label{eq:derivative_matrix}
\end{equation}
and
\begin{equation}
\bD_k^{(t)}
\triangleq
\bH_k^{(t)}
-
\hat{\beta}_k^{(t)}\bE_k^{(t)}.
\label{eq:D_matrix}
\end{equation}



\begin{figure}[!t]
\centering
\includegraphics[width=0.45\textwidth]{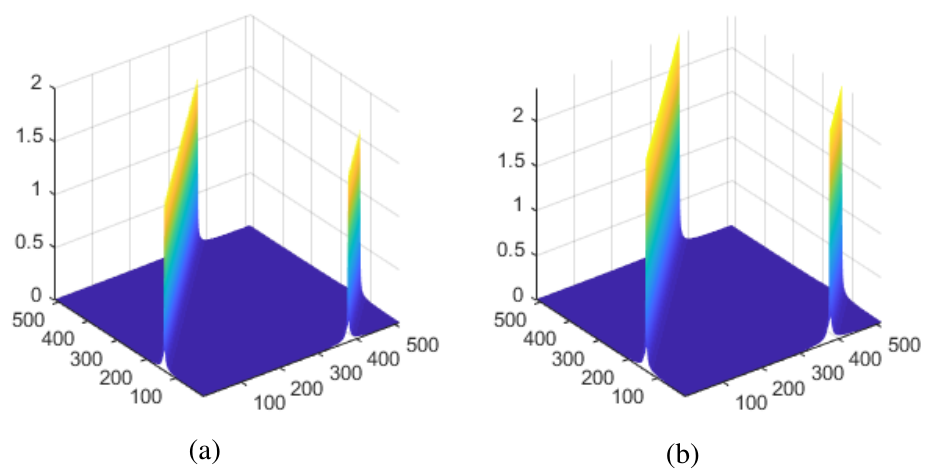}
\centering
\caption{Magnitude patterns of $\bD_k^{(t)}$ and $\bE_k^{(t)}$ for a single path.} 
\label{fig:FigDkEk}
\end{figure}

From \eqref{eq:path_channel_element}, the fractional-Doppler dependence
{is contained in}
\begin{equation}
{\mF}_k(p,q;\beta_k)
=
\sum_{n=0}^{N-1}
\exp\!\left[
-j\frac{2\pi}{N}
n\theta_k(p,q;\beta_k)
\right],
\end{equation}
where
\begin{equation}
\theta_k(p,q;\beta_k)
=
p-q+\alpha_k+\beta_k+2Nc_1l_k .
\end{equation}
Therefore, the $(p,q)$-th entry of the derivative matrix
$\bE_k^{(t)}$ is
\begin{align}
E_k^{(t)}[p,q]
&=
\frac{1}{N}
\exp\!\left\{
j\frac{2\pi}{N}
\left[
Nc_1l_k^2
-ql_k
+Nc_2(q^2-p^2)
\right]
\right\}
\nonumber\\
&\quad\times
\sum_{n=0}^{N-1}
\left(
-j\frac{2\pi n}{N}
\right)
\exp\!\left[
-j\frac{2\pi n}{N}
\theta_k
\!\left(
p,q;\hat{\beta}_k^{(t)}
\right)
\right].
\label{eq:E_element}
\end{align}
As illustrated in Fig.~\ref{fig:FigDkEk}, both $\bD_k^{(t)}$ and $\bE_k^{(t)}$ inherit
the energy concentration of the fractional-Doppler AFDM channel and
exhibit approximately banded structures. This property will later be
exploited to reduce the complexity of the corresponding message-passing
updates.

Substituting \eqref{eq:taylor_channel} into
\eqref{eq:sparse_parametric_model} yields
\begin{align}
\by
&\approx
\sum_{k=1}^{K}
h_k
\left(
\bD_k^{(t)}
+
\beta_k\bE_k^{(t)}
\right)
\bx
+
\bw
\nonumber\\
&=
\sum_{k=1}^{K}
\bD_k^{(t)}
{\bphi}_k
+
\sum_{k=1}^{K}
\bE_k^{(t)}
\boldsymbol{\varphi}_k
+
\bw,
\label{eq:linearized_joint_model}
\end{align}
where
\begin{equation}
\boldsymbol{\phi}_k
\triangleq
h_k\bx,
\qquad
\boldsymbol{\varphi}_k
\triangleq
\beta_k\boldsymbol{\phi}_k .
\end{equation}
For subsequent factor-graph construction, define
\begin{equation}
\bd_k
\triangleq
\bD_k^{(t)}\boldsymbol{\phi}_k,
\qquad
\be_k
\triangleq
\bE_k^{(t)}\boldsymbol{\varphi}_k,
\end{equation}
and
\begin{equation}
\bz_d
\triangleq
\sum_{k=1}^{K}\bd_k,
\qquad
\bz_e
\triangleq
\sum_{k=1}^{K}\be_k .
\end{equation}
The received signal can then be compactly represented as
\begin{equation}
\by
=
\bz_d+\bz_e+\bw.
\label{eq:factor_graph_model}
\end{equation}
Here, $\bz_d$ represents the contribution associated with the locally
linearized nominal channel responses, whereas $\bz_e$ represents the
fractional-Doppler correction component.
The model in \eqref{eq:factor_graph_model} contains an inherent
scale/phase ambiguity between the unknown channel gains and transmitted
symbols. The known pilot symbols at the index set $\mathcal{I}_p$ anchor
this ambiguity, while the remaining symbols at $\mathcal{I}_d$ are
treated as unknown data symbols. Importantly, no pilot guard symbols are introduced; hence, pilot and data observations remain coupled
and are jointly exploited in the Bayesian inference framework developed
in the following subsection.

\begin{figure*}[htbp]
\begin{align}
\frac{\partial \mathcal{F}_k(p,q)}{\partial \beta_k} &= -j2\pi \frac{\exp(-j2\pi\theta)(\exp(-j2\pi\theta/N)-1) - \exp(-j2\pi\theta/N)(\exp(-j2\pi\theta)-1) /N}{(\exp(-j2\pi\theta/N)-1)^2}, \label{eq:partF} 
\end{align}
\end{figure*}

\subsection {Probabilistic Formulation and Factor-Graph Representation}

{We develop a Bayesian method to solve the problem based on the linearized model \eqref{eq:factor_graph_model}. In particular, sparse-Bayesian-learning (SBL) is incorporated into the Bayesian inference method  to jointly estimate the sparse path gains $h_k$, the transmitted symbols, and fractional-Doppler offsets $\beta_k$. SBL automatically identifies active multipath components from the candidate delay–integer-Doppler set via adaptive updates of the precision hyperparameters $\gamma_k$. It is worth mentioning that the channel gain $h_k$ and fractional-Doppler offset $\beta_k$ of the same candidate path are intrinsic properties of a single physical propagation path. Hence, in our design, we make them share a common precision hyperparameter $\gamma_k$, enabling path-granularity sparsity control and ensuring consistent activity determination for both quantities.}

Let $\bz=\bz_d+\bz_e$. Taking advantage of an iterative method, where the estimates in the previous iteration $t$ is available, so the matrices $\{\bD_k^{(t)},\bE_k^{(t)}\}_{k=1}^{K}$ can be treated as known, which are updated after the fractional-Doppler estimates are refined in the current iteration. Based on the current linearization matrices, the joint distribution
can be factorized as
{\begin{align}
&p\!\left(
\by,\bz,\bz_d,\bz_e,
\{\bd_k,\be_k,\boldsymbol{\varphi}_k,\boldsymbol{\phi}_k,
h_k,\beta_k,\gamma_k\}_{k=1}^{K},
\bx
\right)
\nonumber\\
&=
p(\by|\bz)\,
p(\bz|\bz_d,\bz_e)\,
p\!\left(\bz_d|\{\bd_k\}_{k=1}^{K}\right)\,
p\!\left(\bz_e|\{\be_k\}_{k=1}^{K}\right)
\nonumber\\
&\quad\times
\prod\nolimits_{k=1}^{K}
p(\bd_k|\boldsymbol{\phi}_k)\,
p(\be_k|\boldsymbol{\varphi}_k)\,
p(\boldsymbol{\phi}_k|h_k,\bx)\,
p(\boldsymbol{\varphi}_k|\boldsymbol{\phi}_k,\beta_k)
\nonumber\\
&\qquad\qquad\times
p(h_k|\gamma_k)\,
{p(\beta_k)}\,
p(\gamma_k)
\;
p(\bx_d)\,
p(\bx_p)
\nonumber\\
&\triangleq
f_y f_z f_{z_d} f_{z_e}
\prod\nolimits_{k=1}^{K}
f_{d_k}f_{e_k}f_{\phi_k}f_{\varphi_k}
f_{h_k}f_{\beta_k}f_{\gamma_k}
\,f_{x_d}f_{x_p}.
\label{eq:joint_factorization}
\end{align}}

The observation likelihood is
\begin{equation}
f_y(\by,\bz)
=
\mathcal{CN}
\left(
\by;\bz,\lambda^{-1}\bI
\right),
\end{equation}
where $\lambda$ denotes the noise precision. The deterministic
relationships introduced in Section~III-A are represented by
\begin{equation*}
\begin{aligned}
f_z
&=\delta(\bz-\bz_d-\bz_e),
&
f_{z_d}
&=\delta\!\left(\bz_d-\sum_{k=1}^{K}\bd_k\right),
\\
f_{z_e}
&=\delta\!\left(\bz_e-\sum_{k=1}^{K}\be_k\right),
&
f_{d_k}
&=\delta\!\left(\bd_k-\bD_k^{(t)}\boldsymbol{\phi}_k\right),
\\
f_{e_k}
&=\delta\!\left(\be_k-\bE_k^{(t)}\boldsymbol{\varphi}_k\right),
&
f_{\phi_k}
&=\delta\!\left(\boldsymbol{\phi}_k-h_k\bx\right),
\\
f_{\varphi_k}
&=\delta\!\left(\boldsymbol{\varphi}_k-\beta_k\boldsymbol{\phi}_k\right).
\end{aligned}
\end{equation*}
{To exploit the sparse multipath structure, we associate each candidate
delay--integer-Doppler pair with a common SBL precision parameter
$\gamma_k$. Specifically, the path gain is assigned the Gaussian prior
\begin{equation}
f_{h_k}=p(h_k|\gamma_k)
=
\mathcal{CN}
\left(
h_k;0,\gamma_k^{-1}
\right).
\end{equation}
The precision parameter is assigned a Gamma hyperprior,
\begin{equation}
f_{\gamma_k}=p(\gamma_k)
=
\operatorname{Ga}
\left(
\gamma_k;\epsilon,\eta
\right).
\end{equation}
The fractional-Doppler parameter $\beta_k$ is a real scalar confined to $[-0.5,0.5]$, and is assigned an independent uniform prior
\begin{equation}
f_{\beta_k}=p(\beta_k)=\mathcal{U}[-0.5,0.5]. \label{eq:beta_prior}
\end{equation}} 

Let $\mathcal{A}$ denote the QAM constellation with cardinality
$|\mathcal{A}|=M$. For the data symbols,
\begin{equation}
f_{\bx_d}(\bx_d)
=
\prod_{n\in\mathcal{I}_d}
\left[
\frac{1}{M}
\sum_{a\in\mathcal{A}}
\delta(x_n-a)
\right],
\end{equation}
whereas the known pilot symbols are represented by
\begin{equation}
f_{\bx_p}(\bx_p)
=
\prod_{n\in\mathcal{I}_p}
\delta(x_n-\bar{x}_n),
\end{equation}
where $\bar{x}_n$ denotes the known pilot transmitted at index $n$.

\begin{figure*}[!t]
\centering
\includegraphics[width=0.65\textwidth]{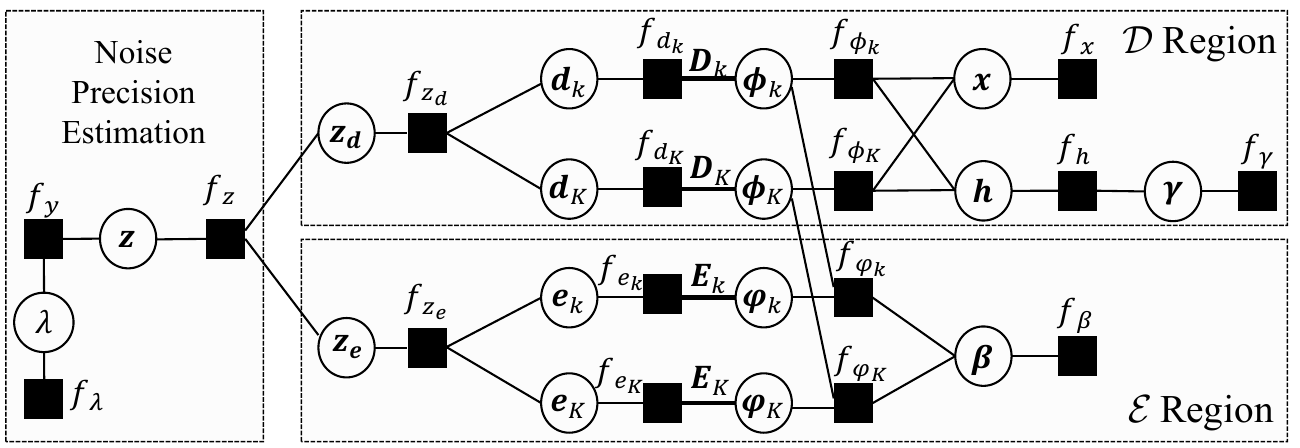}
\centering
\caption{Factor graph of the proposed Bayesian joint channel estimation and symbol detection model.}
\label{fig:FigFactor}
\end{figure*}

The factorization in \eqref{eq:joint_factorization} gives rise to the
factor graph illustrated in Fig.~\ref{fig:FigFactor}. The graph contains two coupled branches corresponding to the
nominal channel contribution $\bz_d$ and the fractional-Doppler
correction $\bz_e$, referred to as the $\mathcal{D}$ and $\mathcal{E}$
regions, respectively. These two regions exchange soft information
through the coupled variables $\boldsymbol{\phi}_k$,
$\boldsymbol{\varphi}_k$, $h_k$, $\beta_k$, and $\bx$. Based on this
factor graph, Bayesian inference aims to obtain the posterior marginals
of the transmitted symbols, path gains, and fractional-Doppler offsets,
e.g., $p(x_n|\by)$, $p(h_k|\by)$, and $p(\beta_k|\by)$. Exact
marginalization is computationally intractable due to the coupled
structure of the model. Therefore, the proposed receiver employs
iterative message passing to approximate these posterior marginals,
thereby enabling joint symbol detection, sparse path recovery, and
fractional-Doppler estimation. The corresponding message-passing
updates are derived in Section~IV.

\section{Proposed Bayesian Message-Passing Receiver}
Based on the factor graph in Fig.~\ref{fig:FigFactor}, we develop an iterative Bayesian
message-passing receiver to approximate the posterior marginals of the
transmitted symbols, channel gains, and fractional-Doppler offsets.
At the $t$-th iteration, the linearization matrices
$\{\bD_k^{(t)},\bE_k^{(t)}\}_{k=1}^{K}$ are treated as fixed and are
updated after refining the fractional-Doppler estimates.

The message-passing procedure consists of three coupled components:
the $\mathcal{E}$ region associated with the fractional-Doppler
correction term $\bz_e$, the $\mathcal{D}$ region associated with the
nominal channel contribution $\bz_d$, and the noise-precision update.
Within the two linear-mixing regions, AMP
is employed to efficiently obtain Gaussian pseudo-observations, while
the multiplicative constraints are handled through Gaussian
moment-matching approximations. The resulting messages are iteratively
exchanged between the two regions until convergence.


\subsection{Message Passing in the $\mathcal{E}$ Region}

The $\mathcal{E}$ region accounts for the fractional-Doppler correction
term
\begin{equation}
\bz_e=\sum_{k=1}^{K}\be_k,\qquad
\be_k=\bE_k^{(t)}\boldsymbol{\varphi}_k,\qquad
\boldsymbol{\varphi}_k=\beta_k\boldsymbol{\phi}_k ,
\label{eq:E_region_model}
\end{equation}
where $\bE_k^{(t)}$ is evaluated at the current fractional-Doppler
estimate $\hat{\beta}_k^{(t)}$. Starting from the observation node,
messages are first propagated through the linear mixing model to obtain
a pseudo-observation of $\boldsymbol{\varphi}_k$. This information is
then combined with the current belief of $\boldsymbol{\phi}_k$ to refine
$\beta_k$, after which updated information is passed back to the
$\mathcal{D}$ region.

\subsubsection{Forward Message Propagation}
Let the incoming message associated with the nominal-channel component
$\bz_d$ be 
\begin{equation}
m_{z_d\to f_z}(\bz_d)
=
\mathcal{CN}\!\left(
\bz_d;
\vec\bz_d,
\vec{\boldsymbol{\nu}}_{z_d}
\right).
\end{equation}
Using $\by=\bz_d+\bz_e+\bw$, the message from $f_z$ to $\bz_e$ is
\begin{equation}
m_{f_z\to z_e}(\bz_e)
=
\mathcal{CN}\!\left(
\bz_e;
\vec{\bz}_e,
\vec{\boldsymbol{\nu}}_{z_e}
\right),
\end{equation}
with
\begin{align}
\vec{\bz}_e
&=
\by-\vec{\bz}_d,
\\
\vec{\boldsymbol{\nu}}_{z_e}
&=
\hat{\lambda}^{-1}\boldsymbol{1}
+
\vec{\boldsymbol{\nu}}_{z_d}.
\label{eq:ze_forward}
\end{align}

Since $\bz_e=\sum_{k=1}^{K}\be_k$, the extrinsic message from
$f_{z_e}$ to $\be_k$ is
\begin{equation}
m_{f_{z_e}\to e_k}(\be_k)
=
\mathcal{CN}\!\left(
\be_k;
\vec{\be}_k,
\vec{\boldsymbol{\nu}}_{e_k}
\right),
\end{equation}
where
\begin{align}
\vec{\be}_k
&=
\vec{\bz}_e
-
\sum_{k'\neq k}
\cev{\be}_{k'},
\\
\vec{\boldsymbol{\nu}}_{e_k}
&=
\vec{\boldsymbol{\nu}}_{z_e}
+
\sum_{k'\neq k}
\cev{\boldsymbol{\nu}}_{e_{k'}} .
\label{eq:ek_extrinsic}
\end{align}
Together with $\be_k=\bE_k^{(t)}\boldsymbol{\varphi}_k$, the above
message induces the linear pseudo-observation model
\begin{equation}
\vec{\be}_k
=
\bE_k^{(t)}\boldsymbol{\varphi}_k
+
\widetilde{\bw}_{e_k}.
\label{eq:E_pseudo_observation}
\end{equation}

The above pseudo-observation model is a typical linear inverse problem, which can be solved by the AMP method with  
a damping mechanism to ensure convergence. 
To obtain a scalar-variance AMP update, the element-wise uncertainty in
$\widetilde{\bw}_{e_k}$ is approximated by
\begin{equation}
\sigma_{e_k}^{2}
=
\left\langle
\vec{\boldsymbol{\nu}}_{e_k}
\right\rangle ,
\end{equation}
where $\langle\cdot\rangle$ denotes averaging over the vector elements. 

The variables $\hat{\bvarphi}_k$, $\bnu_{p_k}^{\varphi}$, and $\bs_k^\varphi$ are taken from the previous iteration. At the first iteration, they are initialized as $\hat{\bvarphi}_k=\boldsymbol{0}_{N\times1}$, $\bnu_{p_k}^{\varphi}=\boldsymbol{1}_{N\times1}$, and $\bs_k^\varphi=\boldsymbol{0}_{N\times1}$. The AMP updates are then performed as follows:


\begin{itemize}
\item Calculate the intermediate variables $\boldsymbol{p}_k^e$ and $\bnu_{p_k}^e$ as
\begin{align}
&\bnu_{p_k}^e = |\bE_k^t|^{.2} \bnu_{\varphi_k} \in \mathbb{R}^{N \times 1}, \quad \forall k \nonumber\\
&\boldsymbol{p}_k^e = \bE_k^t \hat{\bvarphi}_k - \bnu_{p_k}^e \cdot \bs_k^\varphi \in \mathbb{C}^{N \times 1}, \quad \forall k \label{eq:pe}
\end{align}
\item Then calculate the intermediate variables $\bnu_{s_k}^\varphi$ and $\bs_k^\varphi$ as
{\begin{align}
&\bnu_{s_k}^\varphi = 1 ./ \left(\bnu_{p_k}^e + \sigma_{e_k}^2\right) \in \mathbb{R}^{N \times 1}, \quad \forall k, \nonumber\\
&\bs_k^\varphi = \bnu_{s_k}^\varphi \cdot \left(\vec{\be}_k - \boldsymbol{p}_k^e\right) \in \mathbb{C}^{N \times 1}, \quad \forall k \label{eq:se}
\end{align}}
\item Finally, calculate the extrinsic information of message $m_{\varphi_k \to f_{\varphi_k}}(\bvarphi_k) = \mCN(\bvarphi_k; \vec{\bvarphi}_k, \vec{\nu}_{\varphi_k})$, i.e., the extrinsic information of variable $\bvarphi_k$, whose mean and variance are
\begin{align}
&\vec{\bnu}_{\bvarphi_k} = 1 ./ \left(|\bE_k\herm|^{.2} \bnu_{s_k}^\varphi\right) \in \mathbb{R}^{N \times 1}, \nonumber\\
&\vec{\bvarphi}_k = \hat{\bvarphi}_k + \vec{\bnu}_{\varphi_k} \left(\bE_k\herm \bs_k^\varphi\right) \in \mathbb{C}^{N \times 1}, \quad \forall k. \label{eq:qe}
\end{align}
\end{itemize}

Assume that the posterior distribution of variable $\bphi_k$ is $b(\bphi_k) = \mCN(\bphi_k; \hat{\bphi}_k, \bnu_{\phi_k})$, whose update is given in \eqref{eq:bphi}. Using the constraint relationship among variables $\beta_k$, $\bphi_k$, $\bvarphi_k,\forall k$, i.e., $\delta(\bvarphi_k-\beta_k\bphi_k)$, the message $m_{f_{\varphi_k} \to \beta_k}(\beta_k)$ is obtained as
\begin{equation}
m_{f_{\varphi_k} \to \beta_k}(\beta_k) = \mCN\left(\beta_k; \vec{\bbeta}_{k:}, \vec{\bnu}_{\beta_{k:}}\right) \label{eq:fvarp2beta}
\end{equation}
where the mean $\vec{\bbeta}_{k:} \in \mathbb{C}^{N \times 1}$ and variance $\vec{\bnu}_{\beta_{k:}} \in \mathbb{R}^{N \times 1}$ of the message are calculated as
\begin{align}
&\vec{\bbeta}_{k:} = \frac{\hat{\bphi}_k^* \cdot \vec{\bvarphi}_k}{|\hat{\bphi}_k|^2 + \bnu_{\phi_k}}=(\vec{\beta}_{k1},...,\vec{\beta}_{kN})\tra, \nonumber\\
&\vec{\bnu}_{\beta_{k:}} = \frac{\vec{\bnu}_{\varphi_k}}{|\hat{\bphi}_k|^2 + \bnu_{\phi_k}}=(\vec{\nu}_{\beta_{k1}},...,\vec{\nu}_{\beta_{kN}})\tra. \nonumber
\end{align}
{As $\beta_k$ is a physically real scalar, we retain only the real part of the merged mean when aggregating the $N$ element-wise messages into a single scalar message. The corresponding variance and unconstrained mean are given as
\begin{equation}
\nu_{\beta_k} = \left(\sum_n \frac{1}{ \vec{\nu}_{\beta_{kn}}}\right)^{-1} \ \ \text{and} \ \ 
\hat{\beta}_k = \text{Re}\left(\vec{\nu}_{\beta_k} \sum_n \frac{\vec{\beta}_{kn}} { \vec{\nu}_{\beta_{kn}}}\right).
\end{equation}
To enforce the physical constraint \(\beta_k\in[-0.5,0.5]\), the estimated mean is projected onto the feasible interval as $
\hat{\beta}_k
=
\Pi_{[-0.5,0.5]}(\tilde{\beta}_k),
$
where \(\Pi_{[a,b]}(x)=\min\{b,\max\{a,x\}\}\). The resulting Gaussian belief is approximated as
\begin{equation}
b(\beta_k) \simeq  \mathcal{N}\left(\beta_k; \hat{\beta}_k,\nu_{\beta_k}\right). \label{eq:bbeta}
\end{equation}
} 

\subsubsection{Backward Message Propagation}
With the updated posterior statistics of $\beta_k$, the belief of
$\boldsymbol{\varphi}_k=\beta_k\boldsymbol{\phi}_k$ is approximated
by a Gaussian distribution through moment matching:
\begin{equation}
b(\boldsymbol{\varphi}_k)
\simeq
\mathcal{CN}\!\left(
\boldsymbol{\varphi}_k;
\hat{\boldsymbol{\varphi}}_k,
\boldsymbol{\nu}_{\varphi_k}
\right), \label{eq:blf_varphi}
\end{equation}
where
\begin{align}
\hat{\boldsymbol{\varphi}}_k
&=
\hat{\beta}_k^{(t+1)}
\hat{\boldsymbol{\phi}}_k,
\\
\boldsymbol{\nu}_{\varphi_k}
&=
\left|\hat{\beta}_k^{(t+1)}\right|^2
\boldsymbol{\nu}_{\phi_k}
+
\nu_{\beta_k}
|\hat{\boldsymbol{\phi}}_k|^{. 2}
+
\nu_{\beta_k}\boldsymbol{\nu}_{\phi_k}.
\label{eq:varphi_belief}
\end{align}
The corresponding extrinsic message toward
$\boldsymbol{\phi}_k$ is approximated as
\begin{equation}
m_{f_{\varphi_k}\to\phi_k}
(\boldsymbol{\phi}_k)
\simeq
\mathcal{CN}\!\left(
\boldsymbol{\phi}_k;
\widetilde{\boldsymbol{\phi}}_k,
\widetilde{\boldsymbol{\nu}}_{\phi_k}
\right), \label{eq:fvphi2phi}
\end{equation}
with
\begin{align}
\widetilde{\boldsymbol{\phi}}_k
&=
\frac{
\hat{\beta}_k^{(t+1)}
\vec{\boldsymbol{\varphi}}_k
}{
\big|\hat{\beta}_k^{(t+1)}\big|^2
+
\nu_{\beta_k}
},
\\
\widetilde{\boldsymbol{\nu}}_{\phi_k}
&=
\frac{
\vec{\boldsymbol{\nu}}_{\varphi_k}
}{
\big|\hat{\beta}_k^{(t+1)}\big|^2
+
\nu_{\beta_k}
}.
\label{eq:phi_backward}
\end{align}
{The backward Gaussian message associated with each correction component
$\be_k$ is characterized by the AMP output
\begin{equation}
m_{f_{e_k}\to e_k}(\be_k)
=
\mathcal{CN}\!\left(
\be_k;
\bp_k^{e},
\bnu_{p_k}^{e}
\right),
\end{equation}
where $\bp_k^{e}$ and $\bnu_{p_k}^{e}$ are given in \eqref{eq:pe}. }
Consequently, the aggregate message from the $\mathcal{E}$ region to
$\bz_e$ is
\begin{equation}
m_{f_{z_e}\to z_e}(\bz_e)
=
\mathcal{CN}\!\left(
\bz_e;
\cev{\bz}_e,
\cev{\boldsymbol{\nu}}_{z_e}
\right),
\end{equation}
where 
\begin{align}
\cev{\bz}_e
=
\sum_{k=1}^{K}\bp_k^{e}
~~~\text{and }~~~
\cev{\boldsymbol{\nu}}_{z_e}
=
\sum_{k=1}^{K}\bnu_{p,k}^{e}.
\label{eq:ze_backward}
\end{align}
This message is subsequently combined with the observation model to
drive message propagation through the $\mathcal{D}$ region.

\subsection{Message Passing in the $\mathcal{D}$ Region}
The $\mathcal{D}$ region represents the nominal channel contribution, 
\begin{equation}
\bz_d=\sum_{k=1}^{K}\bd_k,\qquad
\bd_k=\bD_k^{(t)}\boldsymbol{\phi}_k,\qquad
\boldsymbol{\phi}_k=h_k\bx .
\label{eq:D_region_model}
\end{equation}
Using the updated information from the $\mathcal{E}$ region, messages
are propagated through the linear mixing model to refine
$\boldsymbol{\phi}_k$. The resulting information is then combined with
the multiplicative constraint $\boldsymbol{\phi}_k=h_k\bx$ to jointly
update the transmitted symbols and sparse path gains.
\subsubsection{Forward Message Propagation}
Let the message provided by the $\mathcal{E}$ region be
\begin{equation}
m_{z_e\to f_z}(\bz_e)
=
\mathcal{CN}\!\left(
\bz_e;
\cev{\bz}_e,
\cev{\boldsymbol{\nu}}_{z_e}
\right).
\end{equation}
Using the observation constraint
$\by=\bz_d+\bz_e+\bw$, the message from $f_z$ to $\bz_d$ is
\begin{equation}
m_{f_z\to z_d}(\bz_d)
=
\mathcal{CN}\!\left(
\bz_d;
\vec{\bz}_d,
\vec{\boldsymbol{\nu}}_{z_d}
\right),
\end{equation}
where
\begin{align}
\vec{\bz}_d
&=
\by-\cev{\bz}_e,
\\
\vec{\boldsymbol{\nu}}_{z_d}
&=
\hat{\lambda}^{-1}\boldsymbol{1}
+
\cev{\boldsymbol{\nu}}_{z_e}.
\label{eq:zd_forward}
\end{align}

Since $\bz_d=\sum_{k=1}^{K}\bd_k$, the extrinsic message from
$f_{z_d}$ to $\bd_k$ is
\begin{equation}
m_{f_{z_d}\to d_k}(\bd_k)
=
\mathcal{CN}\!\left(
\bd_k;
\vec{\bd}_k,
\vec{\boldsymbol{\nu}}_{d_k}
\right), \label{eq:dk_extrinsic}
\end{equation}
with
\begin{align}
\vec{\bd}_k
&=
\vec{\bz}_d
-
\sum_{k'\neq k}
\cev{\bd}_{k'},
\\
\vec{\boldsymbol{\nu}}_{d_k}
&=
\vec{\boldsymbol{\nu}}_{z_d}
+
\sum_{k'\neq k}
\cev{\boldsymbol{\nu}}_{d_{k'}} .
\end{align}
Together with
$\bd_k=\bD_k^{(t)}\boldsymbol{\phi}_k$, the above message induces
the pseudo-observation model
\begin{equation}
\vec{\bd}_k
=
\bD_k^{(t)}\boldsymbol{\phi}_k
+
\widetilde{\bw}_{d_k}.
\label{eq:D_pseudo_observation}
\end{equation}
Equation \eqref{eq:D_pseudo_observation} can also be handled using the AMP. To implement the scalar-variance AMP, the uncertainty of
$\widetilde{\bw}_{d_k}$ is approximated by
\begin{equation}
\sigma_{d_k}^{2}
=
\left\langle
\vec{\boldsymbol{\nu}}_{d_k}
\right\rangle .
\end{equation}

Similar to the previous subsection, the state variables $\hat{\bphi}_k$, $\bnu_{\phi_k}$, and $\bs_k^\phi$ are taken from the previous iteration. At the first iteration, they are initialized as $\hat{\bphi}_k=\boldsymbol{0}_{N\times1}$, $\bnu_{\phi_k}=\boldsymbol{1}_{N\times1}$, and $\bs_k^\phi=\boldsymbol{0}_{N\times1}$. The AMP updates are then performed as follows:


\begin{itemize}
\item  Calculate intermediate variables $\bnu_{p_k}^d$ and $\boldsymbol{p}_k^d$ as
\begin{align}
&\bnu_{p_k}^d = |\bD_k^t|^{.2} \bnu_{\phi_k} \in \mathbb{R}^{N \times 1}, \nonumber\\
& \boldsymbol{p}_k^d = \bD_k^t \hat{\bphi}_k - \bnu_{p_k}^d \cdot \bs_k^\phi \in \mathbb{C}^{N \times 1}, \quad \forall k .\label{eq:pd}
\end{align}

\item Calculate the intermediate variables $\bnu_{s_k}^\phi$ and $\bs_k^\phi$ as
{\begin{align}
&\bnu_{s_k}^\phi = 1 ./ \left(\bnu_{p_k}^d + \sigma_{d_k}^2\right) \in \mathbb{R}^{N \times 1}, \nonumber\\
&\bs_k^\phi = \bnu_{s_k}^\phi \cdot \left(\vec{\bd}_k - \boldsymbol{p}_k^d\right) \in \mathbb{C}^{N \times 1}, \quad \forall k
\end{align}}

\item Calculate the extrinsic information:
\begin{align}
&\vec{\bnu}_{\phi_k} = 1 ./ \left(|\bD_k\herm|^{.2} \bnu_{s_k}^\phi\right) \in \mathbb{R}^{N \times 1}, \nonumber\\
& \vec{\bphi}_k = \hat{\bphi}_k + \vec{\bnu}_{\phi_k}\left(\bD_k\herm \bs_k^\phi\right) \in \mathbb{C}^{N \times 1}, \quad \forall k. \label{eq:qd}
\end{align}
\end{itemize}

Using the cross message $m_{f_{\varphi_k} \to \phi_k}(\bphi_k)$ obtained in \eqref{eq:fvphi2phi}, according to the message passing rule, the message from $\bphi_k$ to $f_{\phi_k}$ is
\begin{equation}
m_{\bphi_k \to f_{\phi_k}}(\bphi_k) = \mCN\left(\bphi_k; \vec{\bphi}_k, \vec{\bnu}_{\phi_k}\right) \label{eq:phi2fphi}
\end{equation}
where the mean and variance are calculated as
\begin{align}
&\vec{\bnu}_{\phi_k} = 1 ./ \left(1 ./ \tilde{\bnu}_{\phi_k} + 1 ./ \bnu_{q_k}^\phi\right) \in \mathbb{R}^{N \times 1},  \nonumber\\
&\vec{\bphi}_k = \vec{\bnu}_{\phi_k} \cdot \left(\tilde{\bphi}_k ./ \tilde{\bnu}_{\phi_k} + \vec{\bphi}_k ./ \bnu_{q_k}^\phi\right) \in \mathbb{C}^{N \times 1}
\end{align}

According to the function node $f_{\phi_k} = \delta(\bphi_k - \bx h_k)$, the message from $f_{\phi_k}, \forall k$ to $\bx$ can be calculated as
\begin{equation}
m_{f_{\phi_k} \to x}(\bx) = \mCN\left(\bx; \vec{\bx}_{k:}, \vec{\bnu}_{x_{k:}}\right) \label{eq:fphi2x}
\end{equation}
where the mean and variance are
\begin{align}
&\vec{\bx}_{k:} = \frac{\vec{\bphi}_k \hat{h}_k^*}{|\hat{h}_k|^2 + \nu_{h_k}}=(\vec{x}_{k1},...,\vec{x}_{kN})\tra \in \mathbb{C}^{N \times 1}, \nonumber\\
&\vec{\bnu}_{x_{k:}} = \frac{\vec{\bnu}_{\phi_k}}{|\hat{h}_k|^2 + \nu_{h_k}}=(\vec{\nu}_{x_{k1}},...,\vec{\nu}_{x_{kN}})\tra \in \mathbb{R}^{N \times 1}. \nonumber
\end{align}

Merging all messages $m_{f_{\phi_k} \to x}(\bx)$ yields the extrinsic information
\begin{equation}
m_{x_n \to f_{x_n}} = \mCN\left(x_n; \vec{x_n}, \vec{\nu}_{x_n}\right), \forall n, \label{eq:x2fx}
\end{equation}
where the mean and variance are
\begin{align}
&\vec{\nu}_{x_n} = 1 ./ \sum\nolimits_{k} 1 ./ \vec{\nu}_{x_{kn}}, \forall n,  \nonumber\\
&\vec{x}_n = \vec{\nu}_{x} \sum\nolimits_{k} \vec{x}_{kn} ./ \vec{\nu}_{x_{kn}}, \forall n.  \nonumber
\end{align}
{As $\bx$ is QAM-modulated, the posterior is calculated as \cite{yuan2022OTFS}, i.e,
\begin{equation}
b(x_n) = \mCN\left(x_n; \hat{x}_n, \nu_{x_n}\right), \forall n. \label{eq:bx}
\end{equation}
with elements defined as
\begin{eqnarray}
\hat{x}_{n} &=& \sum\nolimits_{i=1}^{|\mA|} a_i \bar\xi_{n}^i, \nonumber\\
\nu_{x_{n}} &=& \sum\nolimits_{i=1}^{|\mA|} \bar\xi_{n}^i \left| a_i - \hat{x}_{n} \right|^2. \nonumber
\end{eqnarray}
where
\begin{align}
\bar\xi_{n}^i = \xi_{n}^i / \sum\nolimits_{i'=1}^{|A|} \xi_{n}^{i'}, \ \ \ 
\xi_{n}^i = \exp\left( -\left| a_i - \vec{x}_n \right|^2/\vec{\nu}_{x_n} \right).
\end{align}
In the above equation $a_i$ denotes the $i$-th constellation point in QAM modulation.}

Using the same calculation method as \eqref{eq:fphi2x}, the message from $f_{\phi_k}$ to $h_k$ can be obtained as
\begin{equation}
m_{f_{\phi_k} \to h_k}(h_k) = \mCN\left(h_k; \vec{\bh}_{k:}, \vec{\bnu}_{h_{k:}}\right) \label{eq:fphi2h}
\end{equation}
where the mean and variance are calculated as
\begin{align}
&\vec{\bh}_{k:} = \vec{\bphi}_k \cdot \hat{\bx}^* ./ (|\hat{\bx}|^2 + \bnu_{x})=(\vec{h}_{k1},...,\vec{h}_{kN})\tra \in \mathbb{C}^{N \times 1}, \nonumber\\
&\vec{\bnu}_{h_{k:}} =\vec{\bnu}_{\phi_k}. /  (|\hat{\bx}|^2 + \bnu_{x})=(\vec{\nu}_{h_{k1}},...,\vec{\nu}_{h_{kN}})\tra \in \mathbb{R}^{N \times 1}.\nonumber
\end{align}

Merging the messages $m_{f_{\phi_k} \to h_k}(h_k)$ over all $n$ yields the message from $h_k$ to $f_{h_k}$:
\begin{equation}
m_{h_k \to f_{h_k}}(h_k) = \mCN\left(h_k; \vec{h}_k, \vec{\nu}_{h_k}\right), \quad \forall k
\end{equation}
with mean and variance
\begin{align}
&\vec{\nu}_{h_k} = 1 / \sum\nolimits_{n} 1 / \vec{\nu}_{h_{k,n}},  \vec{h}_k = \vec{\nu}_{h_k} \sum\nolimits_n \vec{h}_{k,n} / \vec{\nu}_{h_{k,n}} \forall k.\nonumber
\end{align}

Then the posterior is obtained as
\begin{equation}
b(h_k) = \mCN\left(h_k; \hat{h}_k, \nu_{h_k}\right), \quad \forall k \label{eq:bh}
\end{equation}
where 
\begin{align}
\hat{h}_k = \frac{\vec{h}_k}{1 + \vec{\nu}_{h_k} \hat{\gamma}_k}, \nu_{h_k} = \frac{\vec{\nu}_{h_k}}{1 + \vec{\nu}_{h_k} \hat{\gamma}_k}, \forall k.
\end{align}

After obtaining the posterior distributions of variables $\beta_k$ and $h_k, \forall k$, referring to the calculation method in \cite{li2026low}, the estimated value of the SBL hyper-prior $\gamma_k, \forall k$ can be obtained as
\begin{equation}
\hat{\gamma}_k = \frac{\epsilon + 1}{\eta + \nu_{h_k} + |\hat{h}_k|^2 }, \quad \forall k. \label{eq:bgamma}
\end{equation}

\subsubsection{Backward Message Propagation}
Using the updated posterior statistics of $h_k$ and $\bx$, the belief
of $\boldsymbol{\phi}_k=h_k\bx$ is approximated by matching its first
two moments,
\begin{equation}
b(\boldsymbol{\phi}_k)
\simeq
\mathcal{CN}\!\left(
\boldsymbol{\phi}_k;
\hat{\boldsymbol{\phi}}_k,
\boldsymbol{\nu}_{\phi_k}
\right), \label{eq:bphi}
\end{equation}
where
\begin{align}
\hat{\boldsymbol{\phi}}_k
&=
\hat{h}_k\hat{\bx},
\\
\boldsymbol{\nu}_{\phi_k}
&=
|\hat{h}_k|^2\boldsymbol{\nu}_x
+
\nu_{h_k}|\hat{\bx}|^{. 2}
+
\nu_{h_k}\boldsymbol{\nu}_x .
\label{eq:phi_belief}
\end{align}
The corresponding backward Gaussian message associated with 
$\bd_k$ is characterized by the AMP output 
\begin{equation}
m_{f_{d_k}\to d_k}(\bd_k)
=
\mathcal{CN}\!\left(
\bd_k;
\bp_k^{d},
\bnu_{p_k}^{d}
\right). \label{eq:fdk2dk}
\end{equation}
where $\bp_k^{d}$ and $\bnu_{p_k}^{d}$ are given in \eqref{eq:pd}. 
Combining the contributions of all candidate paths gives
\begin{equation}
m_{f_{z_d}\to z_d}(\bz_d)
=
\mathcal{CN}\!\left(
\bz_d;
\cev{\bz}_d,
\cev{\boldsymbol{\nu}}_{z_d}
\right),
\end{equation}
where
\begin{align}
\cev{\bz}_d
=
\sum_{k=1}^{K}\bp_k^{d}, ~~~~
\cev{\boldsymbol{\nu}}_{z_d}
=
\sum_{k=1}^{K}\bnu_{p_k}^{d}.
\label{eq:zd_backward}
\end{align}
The resulting message is returned to the observation node and used in
the subsequent $\mathcal{E}$-region update.

\subsection{Noise Precision Estimation} 
After obtaining messages $m_{f_{z_d} \to \bz_d}(\bz_d)$ and $m_{f_{z_e} \to \bz_e}(\bz_e)$, the posterior probability of $\bz$ can be obtained according to the BP message passing algorithm
\begin{equation}
b(\bz) = \mCN\left(\bz; \hat{\bz}, \bnu_{z}\right) \label{eq:bz}
\end{equation}
where the mean and variance are
{\begin{align}
&\bnu_{\bz} = 1 ./ (\hat{\lambda} + \cev{\bnu}_{z_d}^{-1} + \cev{\bnu}_{z_e}^{-1}) \in \mathbb{R}^{N \times 1}, \nonumber\\
&\hat{\bz} = \bnu_{z} \cdot \left((\cev{\bz}_d + \cev{\bz}_e) ./ (\cev{\bnu}_{z_d} + \cev{\bnu}_{z_e}) + \by \hat{\lambda}\right) \in \mathbb{C}^{N \times 1}. \nonumber
\end{align}}
Then the noise precision is calculated as
\begin{equation}
\hat{\lambda}
=
{N}/
{
(\|\by-\hat{\bz}\|_2^2
+
\boldsymbol{1}^{T}\boldsymbol{\nu}_z)
}.
\label{eq:lambda_update}
\end{equation}


The proposed receiver iteratively exchanges soft information between the
$\mathcal{E}$ and $\mathcal{D}$ regions while jointly refining the
fractional-Doppler offsets, sparse path gains, transmitted symbols, SBL
hyperparameters, and noise precision. At each iteration, the
$\mathcal{E}$ region first updates the fractional-Doppler correction
variables and $\{\beta_k\}_{k=1}^{K}$. The resulting messages are then
passed to the $\mathcal{D}$ region, where the transmitted symbols and
channel gains are updated. Subsequently, the SBL hyperparameters and
noise precision are refined, and the updated fractional-Doppler
estimates are used to reconstruct the linearization matrices
$\{\bD_k^{(t+1)},\bE_k^{(t+1)}\}_{k=1}^{K}$ for the next iteration.

\begin{table}[htbp]
\renewcommand{\arraystretch}{1.3}
\caption{Proposed Joint Channel Estimation and Detection Algorithm. }
\label{tab:algorithm}
\centering
\begin{tabular}{l}
\hline
\textbf{Initialization}: \\
$\hat{\bbeta} = \boldsymbol{0}_{K \times 1}$, construct matrices $\bD_k^0$ and $\bE_k^0$  using \eqref{eq:D_matrix} and \eqref{eq:derivative_matrix}; \\
Initialize $\hat{\lambda} = 1$, $\hat{\bvarphi}_k = \boldsymbol{0}_{N \times 1}$, $\bnu_{\varphi_k} = \boldsymbol{1}_{N \times 1}$, $\bs_k^\varphi = \boldsymbol{0}_{N \times 1}$; \\
$\hat{\bgamma} = \boldsymbol{1}_{K \times 1}$, $m_{\bz_d \to f_{z_d}} = \mCN(\bz_d; \boldsymbol{0}_{N \times 1}, \boldsymbol{1}_{N \times 1})$; \\
$m_{\be_k \to f_{z_e}} = \mCN(\be_k; \boldsymbol{0}_{N \times 1}, \boldsymbol{1}_{N \times 1})$, $\hat{\bphi}_k = \boldsymbol{0}_{N \times 1}$, $\bnu_{\phi_k} = \boldsymbol{1}_{N \times 1}$, $\bs_k^\phi = \boldsymbol{0}_{N \times 1}$. \\
\hline
\textbf{For} $it = 1 : {Iter}$ \\
1. Update message $m_{f_z \to z_e}(\bz_e)$ using \eqref{eq:ze_forward}; \\
2. Update messages $m_{f_{z_e} \to e_k}(\be_k), \forall k$ using \eqref{eq:ek_extrinsic}; \\
3. {Update variables $\bnu_{p_k}^e, \boldsymbol{p}_k^e, \bnu_{s_k}^\varphi, \bs_k^\varphi, \vec{\bnu}_{\varphi_k}, \vec{\bvarphi}_k, \forall k$ using \eqref{eq:pe}-\eqref{eq:qe};} \\
4. Update $m_{f_{\varphi_k} \to \beta_k}(\beta_k)$, $m_{\beta_k \to f_{\beta_k}}(\beta_k)$ and $b(\beta_k), \forall k$ using \eqref{eq:fvarp2beta}-\eqref{eq:bbeta}; \\
5. Update $b(\bvarphi_k)$ and $m_{f_{\varphi_k} \to \phi_k}(\bphi_k), \forall k$ using \eqref{eq:blf_varphi} and \eqref{eq:fvphi2phi}; \\
6. Update message $m_{f_{z} \to z_d}(\bz_d)$ using \eqref{eq:zd_forward}; \\
7. Update message $m_{f_{z_d} \to d_k}(\bd_k), \forall k$ using \eqref{eq:dk_extrinsic}; \\
8. {Update variables $\bnu_{p_k}^d, \boldsymbol{p}_k^d, \bnu_{s_k}^\phi, \bs_k^\phi, \vec{\bnu}_{\phi_k}, \vec{\bq}_k, \forall k$ using \eqref{eq:pd}-\eqref{eq:qd};} \\
9. Update messages $m_{\phi_k \to f_{\phi_k}}(\bphi_k), \forall k$ using \eqref{eq:phi2fphi}; \\
10. Update messages $m_{f_{\phi_k} \to x}(\bx), \forall k$ and $m_{x \to f_x}(\bx)$ using \eqref{eq:fphi2x}-\eqref{eq:x2fx}; \\
11. Update $b(\bx)$ and perform symbol decision using \eqref{eq:bx}; \\
12. Update messages $m_{f_{\phi_k} \to h_k}(h_k), \forall k$, $m_{h \to f_h}(\bh)$ and $b(\bh)$ using \eqref{eq:fphi2h}-\eqref{eq:bh}; \\
13. Update hyper-prior $\hat{\bgamma}$ using \eqref{eq:bgamma}; \\
14. Update $b(\bphi_k)$ and $m_{f_{d_k} \to d_k}(\bd_k), \forall k$ using \eqref{eq:bphi} and \eqref{eq:fdk2dk}; \\
15. Update message $m_{f_{z_d} \to z_d}(\bz_d)$ and $b(\bz)$ using \eqref{eq:zd_backward} and \eqref{eq:bz}; \\
16. Update noise precision $\hat{\lambda}$ using \eqref{eq:lambda_update}; \\
17. Update matrices $\bD_k^t$ and $\bE_k^t$ according to $\hat{\bbeta}$ using \eqref{eq:D_matrix} and \eqref{eq:derivative_matrix}. \\
\textbf{End For} \\
\hline
\end{tabular}
\end{table}

\subsection{Complexity Analysis and Comparison }
It can be seen from the algorithm flow that the proposed channel estimation and joint detection method does not require high-complexity operations such as matrix inversion, and all message updates only involve operations such as matrix-vector multiplication. The highest-complexity operations come from updating variables $\bnu_{p_k}^e, \boldsymbol{p}_k^e, \bnu_{s_k}^\varphi, \bs_k^\varphi, \vec{\nu}_{\bvarphi_k}, \vec{\bvarphi}_k, \forall k$ and $\bnu_{p_k}^d, \boldsymbol{p}_k^d, \bnu_{s_k}^\phi, \bq_k, \forall k$, which require $K$ times of matrix-vector multiplication with dimension $N \times N$ and $N \times 1$. Thus, the complexity of the proposed algorithm is $O(K N^2)$. According to \eqref{eq:D_matrix} and Fig. \ref{fig:FigDkEk}, matrices $\bD_k$ and $\bE_k$ are banded sparse matrices, and the number of non-zero elements in each row is $2\kappa_v+1$. Therefore, the complexity of the proposed algorithm can be further reduced to $\mO(KN(2\kappa_v+1))$. 

The grid-refinement (GR) and grid-evolution (GE) algorithms, proposed in \cite{li2026low}, adopted as benchmarks are both built upon the SBL framework. Specifically, the dominant computational complexity of GE-SBL is $O(M_T^3)$, which mainly arises from the inversion of an $M_T$-dimensional covariance matrix. While retaining the same cubic complexity, GR-SBL introduces an additional computational overhead of $O({A}_p M_T^2)$ due to its local grid search. Here $M_T=N_p+Q$ denotes the dimension of the truncated received pilot signal, with $Q$ being the number of guard symbols defined in \eqref{eq:Q}, and ${A}_p$ represents the local search points per path in the GR algorithm. A low-complexity scheme is further proposed in \cite{li2026low}, which reduces the cubic complexity from $O(M_T^3)$ to $O(M_T^3 / C^2)$ by partitioning the observation matrix into $C$ sub-blocks. In contrast, the proposed method in this paper adopts a parametric Bayesian message-passing architecture. It requires no matrix inversion and can fully exploit the banded sparsity of the channel in the DAFT domain to reduce computational load, resulting in significantly lower complexity than the aforementioned channel estimation algorithms.

\section{Numerical Simulation Results}
This section evaluates the performance of the proposed Bayesian joint
channel estimation and symbol detection receiver in representative 5G
high-mobility scenarios, with the main system parameters selected
according to typical settings considered in 3GPP (TR38.901) \cite{3gpp}. Unless
otherwise specified, the number of subcarriers is set to $N=512$, and
QPSK modulation is employed. The carrier frequency and subcarrier
spacing are set to $f_c=10$~GHz and $\Delta f=15$~kHz, respectively,
while the maximum user speed is $500$~km/h. Under the considered
parameter setting, the maximum integer-Doppler index is
$\alpha_{\max}=0$, and the fractional-Doppler offsets are independently
generated according to $\beta_i\sim\mathcal{U}[-0.5,0.5]$.

The number of physical propagation paths, denoted by $P$, ranges from
$1$ to $5$, and the maximum delay index is $l_{\max}=10$. For each
channel realization, the path gains are independently generated as $h_i\sim\mathcal{CN}(0,\eta_i)$,
where the average path power follows an exponentially decaying delay
profile,
\begin{equation}
\eta_k = \exp(-\alpha l_k) / \sum\nolimits_{k'} \exp(-\alpha l_{k'})
\end{equation}
with $\alpha=0.1$. Unless otherwise specified, the effective
half-bandwidth used to retain the significant fractional-Doppler
channel coefficients is set to $\kappa_v=5$.

The proposed Bayesian joint channel estimation and signal detection receiver, named BJCD, is compared with the following benchmark schemes:
GR-SBL and GE-SBL \cite{li2026low}, which perform fractional-Doppler channel estimation
using grid-refinement and grid-evolution SBL, respectively; and an all-pilot benchmark AP, in which all transmitted symbols are assumed
known and only channel-parameter estimation is performed. For symbol
detection, the ideal-CSI detector with the true channel matrix is also
included as a reference, which is denoted as ideal-CSI in the figures.

\begin{table}[htbp]
\centering
\caption{Simulation Parameters}
\label{tab:sim_params}
\begin{tabular}{lll}
\toprule
Parameter & Symbol & Value \\
\midrule
Number of subcarriers & $N$ & 512 \\
Modulation scheme & --- & QPSK \\
Carrier frequency & $f_c$ & 10 GHz \\
Subcarrier spacing & $\Delta f$ & 15 kHz \\
Maximum user speed & $v_{\text{max}}$ & 500 km/h \\
Maximum integer Doppler & $\alpha_{\text{max}}$ & 0 \\
Fractional Doppler & $\beta_k$ & $\text{U}[-0.5, 0.5]$ \\
Number of multipath components & $P$ & 1 $\sim$ 5 \\
Maximum multipath delay & $l_{\text{max}}$ & 10 \\
Channel coefficient distribution & $h_k$ & Gaussian \\
Multipath power decay factor & $\alpha$ & 0.1 \\
Banded channel matrix width & $\kappa_v$ & 5 \\
\bottomrule
\end{tabular}
\vspace{0.5em}
\end{table}


The estimation accuracy is evaluated in terms of the normalized
mean-square errors (NMSEs) of the equivalent channel matrix, fractional
Doppler, and path gains, denoted by $\mathrm{NMSE}_H$,
$\mathrm{NMSE}_\beta$, and $\mathrm{NMSE}_h$, respectively. They are
defined as
\begin{align}
\mathrm{NMSE}_H
&=
10\log_{10}
\frac{\|\bH-\hat{\bH}\|^2}
{\|\bH\|^2},
\label{eq:nmse_H}
\end{align}
where
\begin{equation}
\bH=\sum_{i=1}^{P}h_i\bH_i,
\qquad
\hat{\bH}=\sum_{k=1}^{K}\hat{h}_k\hat{\bH}_k, \nonumber
\end{equation}
and $\hat{\bH}_k$ is constructed using the estimated fractional-Doppler
offset $\hat{\beta}_k$ according to \eqref{eq:daft_received_signal}.
The NMSEs of the fractional-Doppler offsets and path gains are defined as
\begin{align}
\mathrm{NMSE}_\beta
&=
10\log_{10}
\frac{\|\boldsymbol{\beta}-\hat{\boldsymbol{\beta}}\|^2}
{\|\boldsymbol{\beta}\|^2},
\\
\mathrm{NMSE}_h
&=
10\log_{10}
\frac{\|\bh-\hat{\bh}\|^2}
{\|\bh\|^2}. \nonumber
\label{eq:nmse_parameters}
\end{align}

\begin{figure}[!t]
\centering
\includegraphics[width=0.35\textwidth]{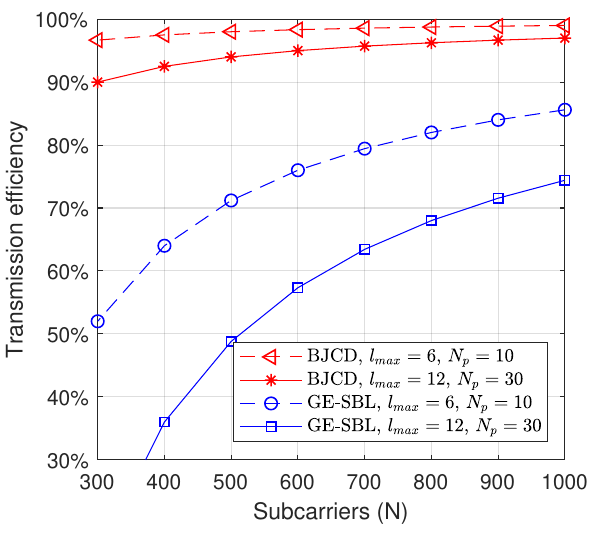}
\centering
\caption{Transmission efficiency versus number of subcarriers and pilot number.} 
\label{fig:FigSpecEff}
\end{figure}

{Fig.~\ref{fig:FigSpecEff} compares the transmission efficiency of the transmission schemes with BJCD and GE-SBL as a function of the number of subcarriers under different maximum-delay $l_{max}$ and pilot length $N_p$. Here, transmission efficiency $\eta$ is defined as the fraction of data-bearing subcarriers in the total subcarriers.  For the proposed BJCD scheme, we have \(\eta = (N - N_p)/N\), and for the benchmark GE-SBL scheme with pilot guard symbols, we have \(\eta = (N - N_p - 2Q)/N\), where \(2Q\) is the number of guard symbols. Since the scheme with GE-SBL requires pilot guard symbols to avoid pilot-data interference, its transmission efficiency can be significantly reduced when the AFDM block length is small or the channel delay spread is large. In contrast, the scheme with BJCD uniformly interleaves pilots with data symbols without pilot guard symbols, and therefore its efficiency loss is determined primarily by the number of pilot symbols. As shown in Fig.~\ref{fig:FigSpecEff}, the scheme with BJCD consistently achieves higher transmission efficiency than the scheme with GE-SBL, with the gain being particularly pronounced for smaller values of $N$.}


\begin{figure}[!t]
\centering
\includegraphics[width=0.35\textwidth]{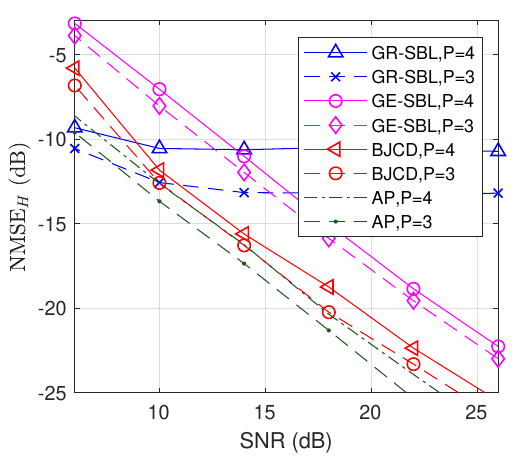}
\centering
\caption{Channel estimation performance versus SNR.} 
\label{fig:FigNMSEHvsSNR}
\end{figure}

\begin{figure}[!t]
\centering
\includegraphics[width=0.5\textwidth]{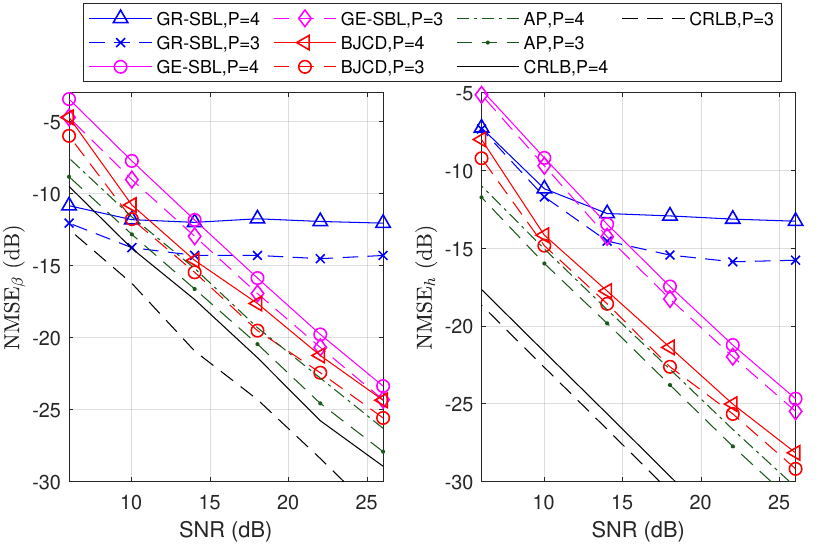}
\centering
\caption{Channel estimation and fractional Doppler estimation performance versus SNR.} 
\label{fig:FigNMSEvsSNR}
\end{figure}

Figs.~\ref{fig:FigNMSEHvsSNR} and \ref{fig:FigNMSEvsSNR} compare the estimation performance of the proposed BJCD receiver with that of GR-SBL and GE-SBL over different SNRs. Specifically, Fig.~\ref{fig:FigNMSEHvsSNR} shows the NMSE of the equivalent channel matrix, $\mathrm{NMSE}_{H}$, while Fig.~\ref{fig:FigNMSEvsSNR} shows the NMSEs of the fractional-Doppler offsets and path gains, $\mathrm{NMSE}_{\beta}$ and $\mathrm{NMSE}_{h}$, respectively. As the SNR increases, all three NMSEs decrease for the considered methods. At a given SNR, BJCD consistently achieves lower $\mathrm{NMSE}_{H}$, $\mathrm{NMSE}_{\beta}$, and $\mathrm{NMSE}_{h}$ than GR-SBL and GE-SBL, with the performance gains becoming more pronounced in the medium-to-high-SNR regime. This improvement mainly results from the joint channel estimation and symbol detection framework, which exploits both pilot observations and soft information from the detected data symbols for channel learning. In addition, the successive Taylor linearization enables continuous refinement of the fractional-Doppler offsets, thereby mitigating the off-grid mismatch. In contrast, GR-SBL and GE-SBL rely on pilot guard symbols and use only pilot observations for channel-parameter estimation, which limits the information available for channel learning.

\begin{figure}[!t]
\centering
\includegraphics[width=0.5\textwidth]{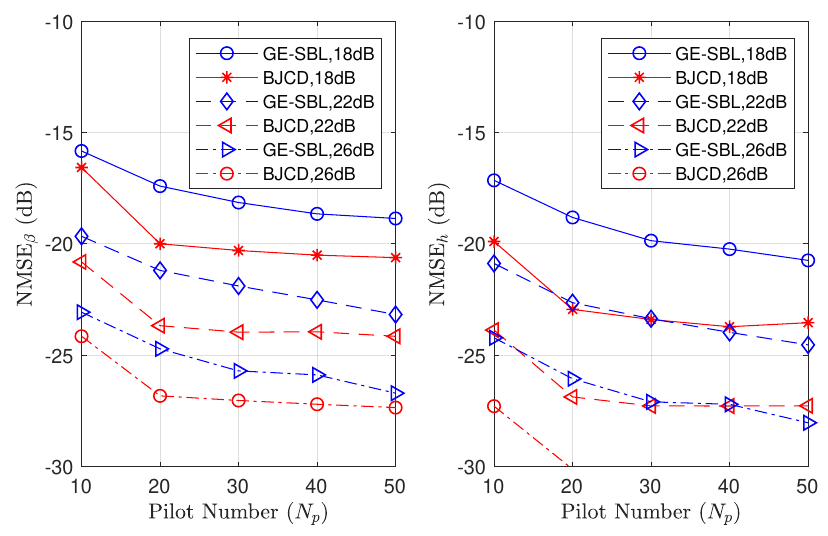}
\centering
\caption{Channel estimation and fractional Doppler estimation performance versus number of pilots.} 
\label{fig:FigNMSEvsPilot}
\end{figure}

Fig.~\ref{fig:FigNMSEvsPilot} compares the fractional-Doppler and channel estimation NMSEs of BJCD and GE-SBL as functions of the number of pilots under different SNRs. As the number of pilots increases, both $\mathrm{NMSE}_{\beta}$ and $\mathrm{NMSE}_{H}$ decrease, since additional pilot observations provide more information for channel-parameter estimation. For a given number of pilots and SNR, BJCD consistently achieves lower estimation errors than GE-SBL, with the performance gain becoming more pronounced at higher SNRs. This improvement results from the joint channel estimation and symbol detection framework, which exploits not only the pilot observations but also soft information from the detected data symbols for channel learning. Consequently, BJCD makes more effective use of the available observations under the same pilot overhead.

\begin{figure}[!t]
\centering
\includegraphics[width=0.5\textwidth]{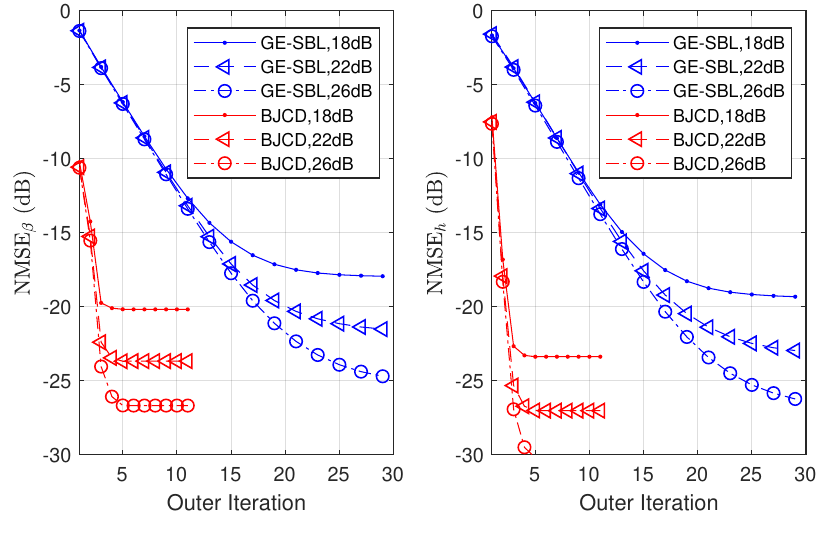}
\centering
\caption{Channel estimation performance versus number of iterations.} 
\label{fig:FigNMSEvsIter}
\end{figure}

Fig.~\ref{fig:FigNMSEvsIter} shows the convergence behavior of the proposed BJCD receiver in terms of the fractional-Doppler and channel estimation NMSEs under different SNRs. As the number of iterations increases, both $\mathrm{NMSE}_{\beta}$ and $\mathrm{NMSE}_{H}$ decrease rapidly and {converge after approximately 5-8 iterations}, with little further improvement thereafter. This indicates that BJCD achieves relatively fast convergence. Moreover, higher SNRs result in lower converged NMSEs as expected.

\begin{figure}[!t]
\centering
\includegraphics[width=0.35\textwidth]{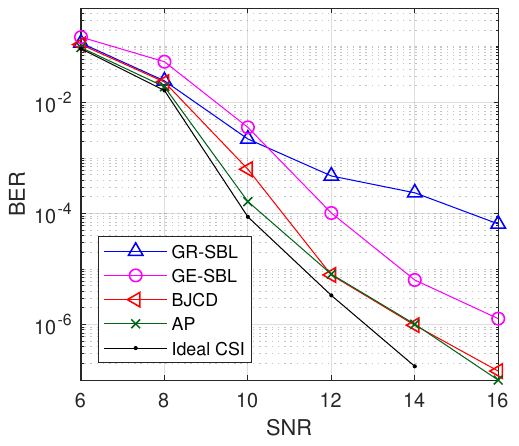}
\centering
\caption{{Signal detection} performance versus SNR.} 
\label{fig:FigBERvsSNR}
\end{figure}

Fig.~\ref{fig:FigBERvsSNR} compares the BER performance of the proposed BJCD receiver with GR-SBL and GE-SBL as channel estimator and EP proposed in \cite{li2026ep} as the detector over different SNRs, with the ideal-CSI detector included as a reference. As the SNR increases, the BERs of all schemes decrease. BJCD consistently achieves lower BER than GR-SBL and GE-SBL and substantially narrows the performance gap to the ideal-CSI detector\cite{li2026ep}. This improvement results from the iterative joint channel estimation and symbol detection framework, in which improved channel estimates enhance symbol detection, while the resulting soft symbol information is subsequently exploited to refine the channel estimates. It is worth highlighting that BJCD achieves reliable symbol detection without requiring pilot guard symbols.

\section{Conclusion}

This paper developed BJCD, a low-complexity Bayesian receiver for spectrally efficient joint channel estimation and symbol detection in fractional-Doppler AFDM systems. By uniformly interleaving pilots with data symbols without pilot guard symbols, BJCD exploits both known pilots and soft data-symbol information for channel learning, thereby avoiding the substantial overhead associated with pilot-isolated transmission. To address fractional Doppler, the AFDM channel was represented by a sparse parametric model over candidate delay--integer-Doppler components, while its nonlinear dependence on the continuous fractional-Doppler offsets was handled through successive first-order Taylor linearization, avoiding exhaustive multidimensional search. A structured factor graph was then constructed to enable iterative Bayesian inference of path activity, channel gains, fractional-Doppler offsets, transmitted symbols, and noise precision within a unified receiver. Numerical results demonstrate that BJCD achieves higher data transmission efficiency and consistently improves equivalent-channel, path-gain, and fractional-Doppler estimation over GR-SBL and GE-SBL, while providing improved symbol-detection performance. 

\appendices
\section{CRLB for Fractional-Doppler and Path-Gain Estimation}
{To provide an oracle benchmark for the estimation performance, we derive the Cramér–Rao lower bounds (CRLBs) for the fractional-Doppler offsets and path gains, assuming that the transmitted symbol vector and the active delay–integer-Doppler support are known.
Let $P$ denote the
number of active paths. The received signal is
\begin{equation}
\by
=
\sum_{k=1}^{P} h_k\bH_k(\beta_k)\bx+\bw,
\end{equation}
where $\bw\sim\mathcal{CN}(\boldsymbol{0},\sigma^2\bI)$,
$h_k\in\mathbb{C}$, and $\beta_k\in(-0.5,0.5)$. The transmitted
symbol vector $\bx$ and noise variance $\sigma^2$ are assumed known.}

Define the equivalent channel response vector and its partial derivative with respect to $\beta_k$,
\begin{equation}
\bu_k(\beta_k) = \bH_k(\beta_k)\bx, \quad 
\dot{\bu}_k(\beta_k) = \frac{\partial \bH_k(\beta_k)}{\partial \beta_k}\bx=\bE_k\bx,
\end{equation}
where $\bE_k$ is defined in \eqref{eq:derivative_matrix}, 
and the cross-correlation terms for arbitrary path pair $k,l=1:P$,
\begin{equation}
a_{kl} = \bu_k^H \bu_l, \quad b_{kl} = \bu_k^H \dot{\bu}_l, \quad c_{kl} = \dot{\bu}_k^H \dot{\bu}_l
\end{equation}

For the parameter vector 
\begin{equation}
\btheta = [\text{Re}(h_1),\text{Im}(h_1),...,\text{Re}(h_P),\text{Im}(h_P),\beta_1,...,\beta_P]^T,  \nonumber
\end{equation}
we have 
\begin{equation}
\ln p(\by|\btheta)=-\frac{1}{\sigma^2}\|\by- \bmu(\btheta)\|_2^2+C \nonumber
\end{equation}
Based on the definition of Fisher information matrix (FIM),
\begin{equation}
J_{mn} = -\mathbb{E}\left[\frac{\partial^2 \ln p}{\partial\theta_m \partial \theta_n}\right]=\frac{2}{\sigma^2}\mathrm{Re}\left\{ \left(\frac{\partial \bu}{\partial \theta_m}\right)^H \frac{\partial \bu}{\partial \theta_n} \right\}
\end{equation}
The overall FIM is partitioned into four sub-blocks according to gain parameters $\bh$ and Doppler parameters $\bbeta$:
\begin{equation}
\bJ = \begin{bmatrix}
\bJ_{hh} & \bJ_{h\beta} \\
\bJ_{\beta h} & \bJ_{\beta\beta}
\end{bmatrix}\in \mathbb{R}^{3P \times 3P}
\end{equation}
where $\bJ_{hh} \in \mathbb{R}^{2P \times 2P}$ corresponds to channel gains, $\bJ_{\beta\beta} \in \mathbb{R}^{P \times P}$ corresponds to Doppler shifts, and $\bJ_{h\beta} \in \mathbb{R}^{2P \times P}$ captures the cross terms, with $\bJ_{\beta h} = \bJ_{h\beta}^T$ by symmetry.

For the complex gain $h_k = \operatorname{Re}(h_k) + j\operatorname{Im}(h_k)$, the partial derivatives of the mean vector are:
\begin{equation}
\frac{\partial \boldsymbol{\mu}}{\partial \operatorname{Re}(h_k)} = \bu_k, \quad \frac{\partial \boldsymbol{\mu}}{\partial \operatorname{Im}(h_k)} = j\bu_k. \nonumber
\end{equation}
For the $(k,l)$-th pair, the corresponding $2 \times 2$ sub-block of $[\bJ_{hh}]_{kl}$ is:

\begin{align}
[\bJ_{hh}]_{kl}
= \frac{2}{\sigma^2}
\begin{bmatrix}
\operatorname{Re}(a_{kl}) & -\operatorname{Im}(a_{kl}) \\
\operatorname{Im}(a_{kl}) & \operatorname{Re}(a_{kl})
\end{bmatrix}.
\end{align}

For any path pair $(k,l)$, the corresponding $2 \times 1$ sub-block of $[\bJ_{h\beta}]_{kl}$ is:
\begin{align}
[\bJ_{h\beta}]_{kl}
= \frac{2}{\sigma^2}
\begin{bmatrix}
\operatorname{Re}(h_l b_{kl})  \\
\operatorname{Im}(h_l b_{kl}) 
\end{bmatrix}.
\end{align}

For any pair of Doppler parameters $(k,l)$:
\begin{align}
\left[\bJ_{\beta\beta}\right]_{kl} &=\mathrm{Re}\left\{ \frac{\partial \bu^H}{\partial \beta_k} \frac{\partial \bu}{\partial \beta_l} \right\}= \frac{2 }{\sigma^2}\mathrm{Re}(h_k^* h_l c_{kl}),
\end{align}


The marginal CRLB for the target parameters with nuisance parameters eliminated is derived via the Schur complement of block Fisher information matrix \cite{Gini2020CRLB}. The partial Fisher information matrix for $\bbeta$ after marginalizing out $\bh$ reads
\begin{equation}
\bJ_\beta = \bJ_{\beta\beta} - \bJ_{\beta h} \bJ_{hh}^{-1} \bJ_{h\beta}.
\end{equation}
Similarly, the marginal FIM of $\bh$ is
\begin{equation}
\bJ_h = \bJ_{hh} - \bJ_{h \beta} \bJ_{\beta\beta}^{-1} \bJ_{\beta h}.
\end{equation}

The Cramér-Rao lower bound matrix for $\bbeta$ and $\bh$ are given by $\bJ_\beta^{-1}$ and $\bJ_h^{-1}$. The diagonal entries correspond to the variance lower bound of each individual Doppler estimate, and the total mean square error lower bounds are
\begin{equation}
\mathrm{CRLB}_\beta = \mathrm{tr}\left( \bJ_\beta^{-1} \right), ~\mathrm{and}~ \mathrm{CRLB}_h = \mathrm{tr}\left( \bJ_h^{-1} \right).
\end{equation}

\bibliographystyle{IEEEtran}
\bibliography{bibliography}

\end{document}